\documentclass[lettersize,journal]{IEEEtran}
\usepackage{color}
\usepackage{bm}
\usepackage{graphicx}
\usepackage{amsmath}
\usepackage{amssymb}
\usepackage[linesnumbered,ruled]{algorithm2e}
\usepackage{multirow}
\usepackage{booktabs}
\usepackage{array}
\usepackage{lipsum}
\usepackage{enumerate}
\usepackage{stfloats}
\usepackage{subfigure}
\usepackage{cases}
\usepackage{diagbox}
\usepackage{cite}
\usepackage{enumitem}
\usepackage{balance}
\usepackage{amsthm}

\theoremstyle{plain}
\newtheorem{defi}{Definition}
\newtheorem{remark}{Remark}

\newtheorem{lemma}{Lemma}

\newtheorem{insight}{Insight}

\allowdisplaybreaks[4]

\begin{document}
	
	\title{Sensing Security Oriented OFDM-ISAC Against Multi-Intercept Threats}
	
	\author{Lingyun Xu, \IEEEmembership{Graduate Student Member~IEEE,}
		Bowen Wang, \IEEEmembership{Graduate Student Member~IEEE,} \\
		Huiyong Li and Ziyang Cheng,~\IEEEmembership{Senior Member~IEEE}
		% \vspace{-2em}
		
		\thanks{
			Manuscript received 28 June 2025; revised 1 March 2026 and 12 May 2026; accepted 22 June 2026.
			This work was supported by the National Natural Science Foundation of China under Grants 62371096 and 62231006, and in part by the Peng Cheng Laboratory Science and Education Foundation - China Mobile Innovation and Technology Fund under Grant XM2603270005.
			(\emph{Corresponding author: Ziyang Cheng})
		}
		\thanks{L. Xu, H. Li, and Z. Cheng are with the School of Information and Communication Engineering, University of Electronic Science and Technology of China, 611731, Chengdu, China. (email: xusherly@std.uestc.edu.cn, \{hyli, zycheng\}@uestc.edu.cn).
			B. Wang is with the Department of Engineering, King's College London, London, WC2R 2LS, UK. (email: bowen.wang@kcl.ac.uk).}}
	
	\maketitle

	\markboth{IEEE Internet of Things Journal}%
	{XU \MakeLowercase{\textit{et al.}}: Sensing Security Oriented OFDM-ISAC Against Multi-Intercept Threats}
	
	\maketitle
	
	\begin{abstract}
		
		In recent years, security has emerged as a critical aspect of integrated sensing and communication (ISAC) systems. While significant research has focused on secure communications, particularly in ensuring physical layer security, the issue of sensing security has received comparatively less attention.
		This paper addresses the sensing security problem in ISAC, particularly under the threat of multi-intercept adversaries. We consider a realistic scenario in which the sensing target is an advanced electronic reconnaissance aircraft capable of employing multiple signal interception techniques, such as power detection (PD) and cyclostationary analysis (CA).
		To evaluate sensing security under such sophisticated threats, we analyze two critical features of the transmitted signal: (i) power distribution and (ii) cyclic spectrum. 
		Further, we introduce a novel ergodic cyclic spectrum metric that leverages the intrinsic mathematical structure of cyclostationary signals to more comprehensively characterize their behavior.
		Building on this analysis, we formulate a new ISAC design problem that explicitly considers sensing security, and we develop a low-complexity, efficient optimization approach to solve it. 
		Simulation results demonstrate that the proposed metric is both effective and insightful, and that our ISAC design significantly enhances sensing security performance in the presence of multi-intercept threats.

	\end{abstract}
	

	\section{Introduction}
	
	\subsection{Context and Motivation}
	\IEEEPARstart{I}{ntegrated} sensing and communication (ISAC) systems support both communication and sensing functionalities within a unified framework by sharing signal processing modules, frequency bands, and hardware resources. This integration significantly improves spectrum efficiency and reduces hardware costs \cite{liu2022integrated,wei2023integrated,cui2021integrating}.
	Thanks to these advantages, ISAC has attracted increasing attention as a key enabler for future 6G applications, including vehicle-to-everything, smart homes, and environmental monitoring \cite{saad2019vision,wang2023road,giordani2020toward,cheng2022lightweight}.
	
	Despite these advancements, security remains a critical and underexplored issue in ISAC systems, stemming from the inherent broadcast nature and spectrum-sharing mechanisms \cite{zou2016survey}. 
	In particular, the dual objectives of communication and sensing introduce unique security challenges from two perspectives: secure communication and secure sensing.
	
	In terms of secure communication within ISAC systems, where the objective is to prevent information leakage and ensure data confidentiality, extensive research efforts have been devoted, particularly in the domains of physical layer security \cite{shiu2011physical,mukherjee2014principles,liu2016physical,wei2022toward,jiang2024physical,chu2023joint,xu2022robust,ren2023robust,wang2024sensing}.
	These studies primarily focus on mitigating the risk of eavesdropping by unauthorized entities. 
	To achieve this, various techniques have been proposed, including artificial noise injection \cite{su2020secure,guo2024secure,liu2022outage}, cooperative jamming \cite{hu2017cooperative,gu2023robust,ye2023robust}, reconfigurable intelligent surface (RIS) assistance \cite{magbool2025survey,kumar2025beamforming,kazymova2025achievable,10643599} and constellation shaping \cite{montezuma2015implementing,bang2020secure,ma2022optimal}, all aimed at degrading the reception quality at potential eavesdroppers while preserving reliable communication for legitimate users.
	
	Sensing security refers to the ability of a system to conduct sensing operations without being easily detected, intercepted, or exploited by adversaries \cite{stove2004low,pace2009detecting}. 
	When sensing security is taken into account in ISAC systems, conventional methods that prioritize sensing accuracy are no longer sufficient.
	Instead, low-probability-of-intercept (LPI) strategies \cite{stove2004low,liu2015lpi,zhang2016lpi} should be exploited to minimize the risk of signal interception while ensuring reliable sensing performance.
	% Sensing security refers to the ability of a system to conduct sensing operations without being easily detected, intercepted, or exploited by adversaries.
	A lack of LPI features in ISAC signals may allow electronic surveillance systems (ESMs) to detect and localize transmission over long distances, thereby compromising the stealth and integrity of the sensing task \cite{stove2004low,pace2009detecting,liu2015lpi,zhang2016lpi}.
	Recent works \cite{magbool2025hiding, han2025sensing, rexhepi2025blinding} have also investigated sensing security in ISAC systems, primarily aiming to prevent unauthorized sensing entities from detecting targets or localizing users through signal or environmental manipulation.
	Compared to communication insecurity, which typically results in data leakage or service disruption, sensing insecurity can entail significantly more severe consequences, including the potential exposure of the ISAC platform’s location, operational status, or strategic intent.

	Although sensing security is critical for real-world ISAC applications, it has received disproportionately less attention compared to its communication counterpart and remains largely underexplored. Motivated by this critical gap, we conduct a comprehensive investigation into sensing security for ISAC, focusing on performance analysis under multi-intercept threats, and beamforming design strategies.

	\subsection{Related Work and Challenges}
	The design of LPI strategies for ISAC systems is closely related to the capabilities of electronic reconnaissance (ER) platforms. 
	In general, ER systems can be categorized based on their interception mechanisms: \textit{(i)} power detection (PD)-based ERs that monitor signal strength, and \textit{(ii)} feature extraction-based ERs that exploit signal structures such as modulation, time-frequency signatures (TFS), or cyclostationary features.
	
	\subsubsection{ISAC Against Power Detection-Based Interception}
	PD-based ER systems detect transmissions by evaluating the received signal's power spectral density. 
	If PD is successful, the ER aircraft can infer not only the presence of the ISAC transmission but also its temporal-spatial activity patterns, thereby increasing the observability of the sensing platform.
	To counter such threats, power-constrained sensing has been proposed, aiming to minimize the radiated power of ISAC signals \cite{shi2018low,shi2019low,wang2020lpi,gong2022joint}.
	For instance, an optimal power allocation framework was proposed in \cite{shi2019low}, where the transmit power across multiple ISAC transmitters was minimized subject to constraints on target detection and communication throughput. 
	Similarly, the authors in \cite{wang2020lpi} designed a power allocation strategy for target time-delay estimation, minimizing radar power consumption while maintaining both estimation accuracy and communication quality-of-service (QoS).
	While reducing transmit power naturally lowers the probability of interception by ER with PD, it also inevitably compromises sensing performance.

	\subsubsection{ISAC Against Feature Extraction-Based Interception}
	More advanced ER systems can perform cyclostationary analysis (CA) or TFS extraction to exploit intrinsic waveform features, even under low received power conditions \cite{gupta2019feature,zilberman2006autonomous,chilukuri2020estimation}.
	If CA is further successful, the ER aircraft can recover waveform structures and system-level parameters, enabling identification of sensing signals.
	To defend against these advanced threats, feature-obfuscated sensing methods have been proposed.
	For example, to combat CA, authors in \cite{liu2023integrated} proposed RIS-aided waveform design, where the Cram{\'e}r-Rao bound for extracting cyclic spectrum elements was minimized subject to radar detection and communication QoS constraints.
	Moreover, authors in \cite{shi2025low} investigated the LPI design for multiple-input multiple-output (MIMO) ISAC systems combating an ER target and multiple jammers, where the transmit waveform and communication precoder were optimized by maximizing radar signal-to-interference-plus-noise ratio (SINR) while guaranteeing communication SINR and suppressing the cyclic frequency sidelobe level.

	Despite several initial attempts toward enabling secure sensing in ISAC systems, multiple technical challenges remain unresolved:
	\begin{itemize}
		\item \textit{Uncertainty in ER decision-making}: Most existing methods rely on fixed or known ER detection decision-making mechanisms, limiting their adaptability and robustness in adversarial or dynamically changing environments.
		\item \textit{Multi-intercept threats}: Modern ER platforms may simultaneously employ multiple intercept strategies, such as PD and CA, necessitating more comprehensive and resilient ISAC design strategies.
	\end{itemize}
	
	These challenges underscore the necessity for a robust, system-level ISAC framework capable of defending against heterogeneous and evolving intercept threats—a gap this pioneering work aims to address.

	\subsection{Contributions}
	The main contributions of this work are summarized as follows.
	\begin{itemize}
		\item \textit{Sensing Security Analysis and Insights:}
		An advanced ER aircraft employing multiple intercept processing methods, i.e., PD and CA, is considered. 
		Under this practical interception setting, the sensing security problem is systematically investigated, where detailed analyses of the intercepted signals are conducted in terms of both received power and cyclic spectrum features. 
		Through this comprehensive analysis, several critical insights are obtained regarding the sensing vulnerability of ISAC signals under multi-intercept threats. 
		Furthermore, by exploring the intrinsic mathematical properties of the ergodic cyclic spectrum matrix, a new design guideline is established to ensure the full emulation of cyclostationary behavior, thereby overcoming the limitations of conventional partial feature-based approaches.

		\item \textit{Robust Security-Aware ISAC System.}
		Since the specific internal decision-making mechanisms by which the ER aircraft integrates information to determine interception are inaccessible, ISAC systems need to possess inherent robustness and security awareness.
		With the sensing security analysis at hand, the ISAC system should: 1) control and reduce the radiated power in the mainlobe region, thereby limiting its electromagnetic visibility; 2) enforce a cyclic spectrum that closely resembles background noise, thereby degrading the effectiveness of distinguishing signal features from noise.
		These capabilities enhance sensing security and robustness of ISAC systems under complex conditions, even when the decision-making mechanisms of the ER aircraft are unknown or potentially evolving.

		\item \textit{Low-Complexity Design and Performance Evaluation.}
		Guided by our analysis and design insights, we formulate a practical beamforming design problem to enable secure sensing in ISAC systems.
		To tackle the high-dimensional non-convex design problem, we propose an effective algorithm to optimize the HBF based on an alternating optimization method.
		Finally, we provide numerical simulations to assess the performance of the proposed design, validating the convergence of the proposed algorithm and the effectiveness in guaranteeing satisfactory communication and sensing performance.
		It also shows the superiority of the proposed LPI-aware secure wideband orthogonal frequency division multiplexing (OFDM)-ISAC system over the conventional schemes in suppressing radiated power and generating noise-like cyclic spectra.
	\end{itemize}
	
	\subsection{Organization and Notations}
	\subsubsection{Organization}
	Section II introduces the system model.
	Section III presents the sensing security performance analysis.
	Section IV formulates the sensing security oriented OFDM-ISAC problem and Section V devises the LPI design against multi-intercept threats and proposes an effective algorithm.
	Section VI demonstrates the numerical simulations.
	Section VII concludes this work.
	
	\subsubsection{Notations} 
	The lower-case boldface letter and upper-case boldface letter denote vectors and matrices, respectively.
	${\bf{A}}(i,j)$ denotes an element in the $i$-th row and the $j$-th column of $\mathbf{A}$. $\mathbb{C}^{n}$ denotes an $n$ dimensional complex-valued vector and $\mathbb{C}^{m\times n}$ denotes an $m$ by $n$ dimensional complex-valued space. $(\cdot)^T$ and $(\cdot)^H$ denote the transpose and conjugate transpose operators, respectively. $\left|  \cdot  \right|$ represents a determinant or absolute value determined by the context. ${\left\|  \cdot  \right\|_F}$ and ${\rm Tr}({\cdot})$ denote the Frobenius norm and trace, respectively. ${\mathbb E}\left\{   \cdot \right\}$ denotes expectation. $\Re \left\{  \cdot  \right\}$  denotes the real part of a complex-valued number. $\mathcal{C}\mathcal{N}\left( {0,{\mathbf{R}}} \right)$ denotes the complex Gaussian distribution with zero mean and covariance ${\mathbf{R}}$.
	${\rm{Diag}}\left(\cdot \right)$ denotes forming diagonal matrix from the input.

	\section{System Model}
	
	\begin{figure*}[!t]
		\centering  \includegraphics[width=1\linewidth]{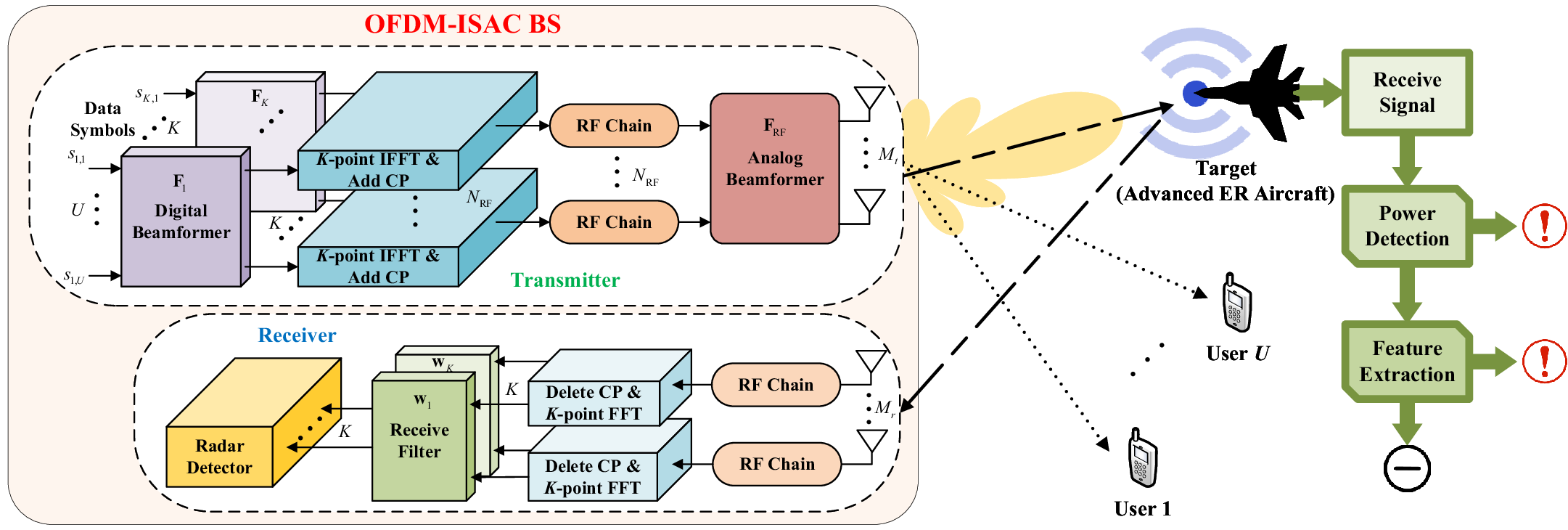}
		\vspace{-1.5em}
		\caption{An illustration of an OFDM-ISAC system with hybrid beamforming architecture at the transmitter in the presence of an advanced ER aircraft.}
		\label{fig:scene_graph}
		\vspace{-1em}
	\end{figure*}
	
	As shown in Fig. \ref{fig:scene_graph}, we consider a wideband OFDM-ISAC base station (BS) with $K$ subcarriers, which serves $U$ single-antenna communication users while simultaneously probing a single-antenna target of interest.
	Specifically, the transmitter is equipped with $M_t$ antennas and the colocated radar receiver is equipped with $M_r$ receive antennas \footnote{In this work, the considered OFDM-ISAC system operates in a time-division manner, where signal transmission and echo reception are separated in the time domain \cite{liu2022integrated}. As a result, transmitter-to-receiver leakage and the associated self-interference issue are inherently avoided and thus not considered in the system model.}. They both adopt uniform linear arrays (ULAs), with the element spacing $d$.
	
	\subsection{Transmission Model}
	
	To cut down the tremendous power consumption and hardware overhead introduced by massive MIMO techniques, the transmitter adopts a hybrid beamforming (HBF) architecture with $N_{\rm RF}$ RF chains \cite{alkhateeb2014channel,11363235,wang2022partially}.
	The transmit data symbol at the $k$-th subcarrier is denoted by ${{\bf{s}}_k} = \left[ {{s_{k,1}}, \ldots ,{s_{k,U}}} \right]^T \in {{\mathbb C}^U}$, assuming ${\mathbb E}\left\{ {{{\bf{s}}_k}{\bf{s}}_k^H} \right\} = {{\bf{I}}_U}$.
	The data symbol is firstly processed by the digital beamformer ${{\bf{F}}_k} = \left[ {{{\bf{f}}_{k,1}}, \ldots ,{{\bf{f}}_{k,U}}} \right] \in {{\mathbb C}^{{N_{\rm RF}} \times U}}$ in the frequency domain.
	Then, the signals are converted to the time domain by $N_{\rm RF}$ $K$-point inverse fast Fourier transforms (IFFTs) and added with the cyclic prefix (CP).
	After the up-conversion to the radio frequency (RF) domain by $N_{\rm RF}$ RF chains, the analog signals are processed by the analog beamformer ${{\bf{F}}_{{\rm{RF}}}} \in {{\mathbb C}^{{M_t} \times {N_{\rm RF}}}}$ implemented by phase shifters (PSs), i.e., $\left| {{{\bf{F}}_{{\rm{RF}}}}\left[ {i,j} \right]} \right| = 1,\forall i,j$.
	Finally, the ISAC signals are emitted by $M_t$ antennas.
	Therefore, the transmitted signal from the OFDM-ISAC BS at the time instant $t\in [0,\Delta t]$ is given by
	\begin{equation}\label{eq:1}
		{\bf{x}}\left( t \right) = {{\bf{F}}_{{\rm{RF}}}}\sum\limits_{k = 1}^K {{{\bf{F}}_k}{{\bf{s}}_k}{e^{\jmath 2\pi {f_k}t}}},
	\end{equation}
	where ${f_k} = {f_c} + \left( {k - \frac{K}{2}} \right)\Delta f$ denotes the frequency at the $k$-th subcarrier, with the center frequency $f_{\rm c}$, the subcarrier spacing $\Delta f = \frac{B}{K} = \frac{1}{\Delta t}$, and the system bandwidth $B$.

	Furthermore, we examine the impact of the CP on the sensing range.
	In CP-OFDM systems, a CP of $N_{\rm cp}$ is appended to mitigate inter-symbol interference \cite{sturm2009novel,liu2019evaluation,liu2020time}.
	From a sensing perspective, the CP duration plays a critical role in determining the maximum observable target delay.
	When the echo delay exceeds the CP length, inter-symbol interference arises, leading to SINR degradation and impaired sensing performance.
	To ensure ISI-free sensing, we assume that the target delays lie within the CP-protected interval. 
	Accordingly, the maximum sensing range is upper bounded by $R_{\rm max,cp} = v\frac{N_{\rm cp}}{2B}$, with system bandwidth $B$ and electromagnetic wave propagation speed $v$.
	
	\subsection{Reception Model}

	\subsubsection{Communication Reception Model}
	
	For the downlink communication, the received signal at the $u$-th user at the $k$-th subcarrier is given by
	\begin{equation}\label{eq:2}
		{r_{k,u}} = {\bf{h}}_{k,u}^H{{\bf{F}}_{{\rm{RF}}}}{{\bf{f}}_{k,u}}{{\bf{s}}_k}
		+ \sum\limits_{i \ne u}^U {{\bf{h}}_{k,u}^H{{\bf{F}}_{{\rm{RF}}}}{{\bf{f}}_{k,i}}{{\bf{s}}_k}}  + {n_{k,u}},
	\end{equation}
	where ${n_{k,u}}$ denotes the complex additive white Gaussian noise (AWGN) with power spectral density $\sigma_{\rm C}^2$, ${\bf h}_{k,u}\in {\mathbb C}^{M_t}$ denotes the channel state information (CSI) vector from the transmitter to the $u$-th user at the $k$-th subcarrier \cite{10858124,cheng2021hybrid,li2020dynamic}.
	
	%Based on \eqref{eq:2}, the achievable communication rate for the $u$-th user at the $k$-th subcarrier is 
	%\begin{equation}\label{eq:3}
	%{\rm{R}}_{k,u}\left( {{{\bf{F}}_{{\rm{RF}}}},{{\bf{F}}_k}} \right) = {\rm{log}}_2\left(1 + {\rm{SIN}}{{\rm{R}}_{k,u}}\left( {{{\bf{F}}_{{\rm{RF}}}},{{\bf{F}}_k}} \right) \right),
	%\end{equation}
	%with the signal-to-interference-plus-noise ratio (SINR) of the received signal
	%\begin{equation}\label{eq:4}
	%{\rm{SIN}}{{\rm{R}}_{k,u}}\left( {{{\bf{F}}_{{\rm{RF}}}},{{\bf{F}}_k}} \right) = \frac{{{{\left| {{\bf{h}}_{k,u}^H{{\bf{F}}_{{\rm{RF}}}}{{\bf{f}}_{k,u}}} \right|}^2}}}{{\sum\limits_{i \ne u}^U {{{\left| {{\bf{h}}_{k,u}^H{{\bf{F}}_{{\rm{RF}}}}{{\bf{f}}_{k,i}}} \right|}^2}}  + \sigma _{\rm{C}}^2}}.
	%\end{equation}
	
	\subsubsection{Radar Reception Model}
	
	In this paper, the radar receiver co-located with the transmitter needs to collect the echo signals in order to detect the target of interest
	at the angle $\theta_{E}$ in the presence of $I$ clutter sources $\theta_{i,k}, \forall i,k$.
	The received signal is down-converted via $M_r$ RF chains.
	After removing the CPs and transforming the time-domain signal into the frequency domain via $M_r$ $K$-point fast Fourier transforms (FFTs), which can be expressed as
	\begin{equation}
		\begin{aligned}
			&{{\bf{y}}_{{\rm{R}},k}} = \underbrace {{\varsigma _{{\rm{E}},k}}{{\bf{a}}_r}\left( {{\theta _{\rm{E}}},{f_k}} \right){\bf{a}}_t^H\left( {{\theta _{\rm{E}}},{f_k}} \right){{\bf{F}}_{{\rm{RF}}}}{{\bf{F}}_k}{{\bf{s}}_k}}_{{\rm{signal}}} \\
			&\qquad\;\;\;+ \underbrace {\sum\limits_{i = 1}^I {{\varsigma _{i,k}}{{\bf{a}}_r}\left( {{\theta _{i,k}},{f_k}} \right){\bf{a}}_t^H\left( {{\theta _i},{f_k}} \right){{\bf{F}}_{{\rm{RF}}}}{{\bf{F}}_k}{{\bf{s}}_k}} }_{{\rm{clutters}}} + \underbrace {{{\bf{z}}_{{\rm{R}},k}}}_{{\rm{noise}}},
		\end{aligned}\label{eq:3}
	\end{equation}
	where ${\varsigma _{{{\rm{E}},k}}} \sim {\mathcal {CN}}(0, \sigma_{{\rm E},k}^2)$ and ${{\varsigma _{i,k}}} \sim {\mathcal {CN}}(0, \sigma_{i,k}^2)$ denote the complex amplitude coefficients including the path loss and the radar cross section (RCS) corresponding to the ER aircraft and the $i$-th clutter at the $k$-th subcarrier, respectively.
	${{\mathbf{z}}_{{\rm{R}},k}}\sim\mathcal{C}\mathcal{N}\left( {0,\sigma _{\rm{R}}^2{{\mathbf{I}}_{{M_r}}}} \right)$ denotes the AWGN.
	Besides, ${{\bf{a}}_t}\left( {\theta ,f} \right) = \frac{1}{{\sqrt {{M_t}} }}{[ {{e^{\jmath 2\pi f{\tau _1}}},{e^{\jmath 2\pi f{\tau _2}}}, \ldots ,{e^{\jmath 2\pi f{\tau _{{M_t}}}}}} ]^T} \in {{\mathbb C}^{{M_t}}}$ and ${{\bf{a}}_r}\left( {\theta ,f} \right) = \frac{1}{{\sqrt {{M_r}} }}{[ {{e^{\jmath 2\pi f{\tau _1}}},{e^{\jmath 2\pi f{\tau _2}}}, \ldots ,{e^{\jmath 2\pi f{\tau _{{M_r}}}}}} ]^T} \in {{\mathbb C}^{{M_r}}}$ denote the space-frequency transmit and receive steering vectors, respectively,
	$\tau_m = {{(m - 1)d\sin \theta }}/{v}$.
	Unlike the narrowband approximation ${f_k} \approx {f_c}$, the proposed steering vectors preserve the frequency dependence.
	As a result, the beam squint effect caused by the wideband transmission is inherently reflected in the equation \eqref{eq:3}.

	Since the receiver has no prior knowledge of the ER aircraft location, it applies a spatial matched filter based on the steering vector, ${{\bf{w}}_k} = {{\bf{a}}_r}\left( {\theta_{\rm scan} ,{f_k}} \right) \in {{\mathbb C}^{{M_r}}}, \forall k$, where the scanning angle $\theta_{\rm scan} \in [-90^\circ,90^\circ]$.
	By scanning the entire angular domain, the maximum output SINR is attained when $\theta_{\rm scan}$ coincides with the actual direction of the ER aircraft, i.e., $\theta_{\rm scan} = \theta_{\rm E}$.
	Therefore, without loss of generality, we consider ${\bf w}_k = {\bf a}_r(\theta_{\rm E}, f_k), \forall k$, and the corresponding filter output can be expressed as
	\begin{equation}
		\begin{aligned}
			{{\tilde y}_{{\rm{R,}}k}} = &\varsigma _{{\rm{E}},k}{\bf{a}}_t^H\left( {{\theta _{\rm{E}}},{f_k}} \right){{\bf{F}}_{{\rm{RF}}}}{{\bf{F}}_k}{{\bf{s}}_k} \\
			&+ \sum\limits_{i = 1}^I {{\varsigma _{i,k}}{\bf{a}}_r^H\left( {{\theta _{\rm{E}}},{f_k}} \right){{\bf{a}}_r}\left( {{\theta _i},{f_k}} \right){\bf{a}}_t^H\left( {{\theta _i},{f_k}} \right){{\bf{F}}_{{\rm{RF}}}}{{\bf{F}}_k}{{\bf{s}}_k}} \\
			&+ {\bf{a}}_r^H\left( {{\theta _{\rm{E}}},{f_k}} \right){{\bf{z}}_{{\rm{R}},k}}.
		\end{aligned}\label{eq:4}
	\end{equation}

	\subsubsection{ER Aircraft Reception Model}
	
	Suppose there exists an advanced ER aircraft that employs the broad-beam and wide-band reception modes to passively receive the surrounding signals.
	The signal received by the advanced ER aircraft is given by
	\begin{equation}\label{eq:5}
		{\bf{r}} = \sum\limits_{k = 1}^K {{\beta _k}{\bf{a}}_t^H\left( {{\theta _{\rm{E}}},{f_k}} \right){{\bf{F}}_{{\rm{RF}}}}{{\bf{F}}_k}{{\bf{s}}_k}{\bf{d}}\left( {{f_k}} \right)}  + {\bf{z}},
	\end{equation}
	where ${\bf{z}}\sim{\cal C}{\cal N}\left( {0,\sigma _{\rm{E}}^2{{\bf{I}}_N}} \right)$ denotes the AWGN, ${\bf{d}}\left( {{f_k}} \right) = {\left[ {1,{e^{\jmath 2\pi {f_k}{T_s}}}, \ldots ,{e^{\jmath 2\pi {f_k}\left( {N - 1} \right){T_s}}}} \right]^T} \in {{\mathbb C}^N}$, with the number of sampling points $N$ and sampling period $T_s = \frac{1}{N \Delta f}$. 
	Besides, ${\beta _k}$ denotes the complex channel coefficient for the $k$-th subcarrier, capturing both the path loss and the phase shift due to propagation delay.
	
	\section{Sensing Security Performance Analysis}

	In this section, we focus on the analysis of sensing security performance.
	We first analyze the properties of intercepted signals and then exploit the properties to obtain a feasible LPI design guideline.
	
	Before starting our analysis, it is essential to first clearly introduce the sensing security challenges in the presence of the advanced ER aircraft.
	As shown in Fig. \ref{fig:scene_graph}, in this paper, we assume that the advanced ER aircraft can scan and receive signals emitted from the OFDM-ISAC BS and then perform multiple intercept processing methods:
	\textit{1) PD:} By measuring whether the power of the received signal exceeds a preset threshold, the ER aircraft can determine the existence of radar signals \cite{pace2009detecting}. 
	\textit{2) CA:} By analyzing the cyclostationary of the received signals, the ER aircraft can distinguish the radar signals from the background noise \cite{pace2009detecting}.
	Such a complex scenario significantly complicates the task of ensuring sensing security in ISAC systems, as the sensing signals are exposed to both power-based and feature-based detection techniques.
	These advanced intercept threats make it difficult to maintain sensing security through traditional waveform design or power control alone, and necessitate new approaches that address both signal power and feature observability.

	It should be noted that, to intercept the transmitted signal from the BS, the ER aircraft is assumed to possess prior knowledge of several long-term system parameters, including the center frequency $f_c$, the number of subcarriers $K$, and the subcarrier spacing $\Delta f$. 
	Such information is required for configuring the frequency indices in cyclostationary spectrum computation and is commonly assumed in worst-case interception analysis \cite{liu2023integrated,shi2025low}.
	
	\subsection{Properties of Intercepted Signals}
	
	To evaluate the sensing security performance, the power and cyclostationary behavior of the received signals should be analyzed.
	Based on the received signal \eqref{eq:5}, the power collected by the advanced ER aircraft across $K$ subcarriers can be directly given by
	\begin{equation}\label{eq:6}
		{P_{\rm{E}}} = \mathbb{E}\left\{ {\left\| {\mathbf{r}} \right\|_F^2} \right\}
		=\sum\limits_{k = 1}^K {\left\| {{\beta _k}{\bf{a}}_t^H\left( {{\theta _{\rm{E}}},{f_k}} \right){{\bf{F}}_{{\rm{RF}}}}{{\bf{F}}_k}} \right\|_F^2}  + K\sigma _{\rm{E}}^2.
	\end{equation}
	
	To further study the cyclostationary property, the signals are converted into the frequency domain by the $N$-point FFTs, which can be expressed as
	\begin{equation}\label{eq:7}
		{{{\bf{\mathord{\buildrel{\lower3pt\hbox{$\scriptscriptstyle\frown$}} 
							\over r} }}}} = {\rm{FFT}}\left( {{{\bf{r}}}} \right) = {{\bf{V}}_{{\rm{FFT}}}}{{\bf{r}}} \in {{\mathbb C}^N},
	\end{equation}
	where ${{\bf{V}}_{{\rm{FFT}}}} \in {{\mathbb C}^{N \times N}}$ denotes the FFT matrix defined by
	\begin{equation}\label{eq:8}
		{{\bf{V}}_{{\rm{FFT}}}}\left[ {i,j} \right] = \exp \left( { - \jmath \frac{{2\pi }}{N}\left( {i - 1} \right)\left( {j - 1} \right)} \right),i,j \in \left[ {1,N} \right].
	\end{equation}
	
	Based on the frequency-domain signal \eqref{eq:7}, we can derive its corresponding cyclic spectrum \cite{gardner1986spectral} by
	
	\begin{align}
		&{{\bf{C}}_{\rm{E}}}\left( {m,n} \right) = \frac{1}{W}\sum\limits_{w =  - {{W \over 2}}}^{{{W \over 2}} - 1} {{\bf{\mathord{\buildrel{\lower3pt\hbox{$\scriptscriptstyle\frown$}} 
						\over r} }}\left( {m + \frac{n}{2} + w} \right){{{\bf{\mathord{\buildrel{\lower3pt\hbox{$\scriptscriptstyle\frown$}} 
								\over r} }}}^*}\left( {m - \frac{n}{2} + w} \right)} , \nonumber\\
		&m \in {\cal M},n \in {\cal N}, \label{eq:9}
	\end{align}
	where $W$ denotes the length of analysis window of the advanced ER aircraft, and ${\cal M} = \left\{ {0,1,2, \ldots ,N - W} \right\}$ and ${\cal N} = \left\{ {0,2,4, \ldots ,N - W} \right\}$.
	Besides, we define the cyclic frequency and frequency as $\alpha  = n{f_s}$ and $f = m{f_s}$, respectively, with the sampling frequency $f_s$.
	
	Substituting \eqref{eq:7} into \eqref{eq:9}, we can further rewrite the cyclic spectrum of the received signal as 
	\begin{equation}\label{eq:10}
		{{\bf{C}}_{\rm{E}}}\left( {m,n} \right) = {{{\bf{\mathord{\buildrel{\lower3pt\hbox{$\scriptscriptstyle\frown$}} 
							\over r} }}}^H}{{\bf{\Lambda }}_{m,n}}{\bf{\mathord{\buildrel{\lower3pt\hbox{$\scriptscriptstyle\frown$}} 
					\over r} }},m \in {\cal M},n \in {\cal N},
	\end{equation}
	where we define 
	\begin{equation}
		{{\bf{\Lambda }}_{m,n}} = \frac{1}{W}\sum\limits_{w =  - {{W \over 2}}}^{{ W \over 2} - 1} {{\bf{J}}_{m - {n \over 2} + w}^H{{\bf{e}}_1}{\bf{e}}_1^T{{\bf{J}}_{m + {n \over 2} + w}}},\nonumber
	\end{equation}
	${{\bf{J}}_l}\left[ {i,j} \right] = \left\{ \begin{array}{l}
		1,\;i = j + l\\
		0,\;{\rm{otherwise}}
	\end{array} \right.$, and ${{\bf{e}}_1} = {\left[ {1_{(1)},0, \ldots ,0} \right]^T}$.
	
	Note that the cyclic spectrum ${\bf C}_{\rm E}$ exhibits randomness due to the randomness of the received signal \eqref{eq:10}.
	To have a better analysis of the characteristics of the cyclic spectrum, we give the following definition.
	\begin{defi}{(Ergodic Cyclic Spectrum)}\label{def:1}:
		To analyze the characteristics of the random cyclic spectrum \eqref{eq:10}, we introduce an ergodic cyclic spectrum, ${{\bf{\tilde C}}_{\rm{E}}}$, which is defined by
		\begin{equation}\label{eq:11}
			{{\bf{\tilde C}}_{\rm{E}}}\left( {m,n} \right) = {\mathbb E}\left\{ {{{\bf{C}}_{\rm{E}}}\left( {m,n} \right)} \right\},m \in {\cal M},n \in {\cal N}.
		\end{equation}
		Based on \eqref{eq:10}, we notice that ${{{\bf{\Lambda }}_{m,n}}}, m \in {\cal M},n \in {\cal N}$ are deterministic and ${{\bf{\mathord{\buildrel{\lower3pt\hbox{$\scriptscriptstyle\frown$}} \over r} }}}$ is random. 
		To this end, the ergodic cyclic spectrum \eqref{eq:11} can be further rewritten as
		\begin{equation}\label{eq:12}
			\begin{aligned}
				&{{{\bf{\tilde C}}}_{\rm{E}}}\left( {m,n} \right) = {\rm{Tr}}\left\{ {{\mathbb E}\left\{ {{\bf{r}}{{\bf{r}}^H}} \right\}{\bf{V}}_{{\rm{FFT}}}^H{{\bf{\Lambda }}_{m,n}}{{\bf{V}}_{{\rm{FFT}}}}} \right\},\\
				&m \in {\cal M},n \in {\cal N}.
			\end{aligned}
		\end{equation}
	\end{defi}
	
	Based on \textbf{Definition \ref{def:1}}, the ergodic cyclic spectrum of the received signal can be expressed by
	\begin{equation}\label{eq:13}
		{{\bf{\tilde C}}_{\rm{E}}}\left( {m,n} \right) = {\rm{Tr}}\left\{ {\left( {\bf{R}}_\Xi  + {\bf{\Omega }} \right){\bf{V}}_{{\rm{FFT}}}^H{{\bf{\Lambda }}_{m,n}}{{\bf{V}}_{{\rm{FFT}}}}} \right\},
	\end{equation}
	where ${{\bf{R}}_\Xi } \!=\! {\bf{D\Xi }}{{\bf{D}}^H} \! \in \! {\mathbb C}^{N \times N}$, ${\bf{D}} \!=\! [ {\bf{d}}\left( {{f_1}} \right), \ldots , {\bf{d}}\left( {{f_K}} \right) ] \in {{\mathbb C}^{N \times K}}$ and ${\bf{\Xi }} = {\rm{Diag}}( {{| {{\beta _1}} |}^2}\| {{\bf{a}}_t^H\left( {{\theta _{\rm{E}}},{f_1}} \right){{\bf{F}}_{{\rm{RF}}}}{{\bf{F}}_1}} \|_F^2,$   $ \ldots ,{{| {{\beta _K}} |}^2}\| {{\bf{a}}_t^H\left( {{\theta _{\rm{E}}},{f_K}} \right){{\bf{F}}_{{\rm{RF}}}}{{\bf{F}}_K}} \|_F^2 ) \! \in \! {{\mathbb C}^{K \times K}}$,
	${\bf{\Omega }} = {\rm{Diag}}( \sigma _{\rm{E}}^2, $  $\ldots ,\sigma _{\rm{E}}^2 ) \in {{\mathbb C}^{N \times N }}$.
	
	Having examined the fundamental properties of intercepted signals, it is important to highlight their practical implications and differences from classical covert communication. These points are summarized in the following remark.
	\begin{remark}(Implications of Interception)
		If the ER aircraft successfully performs PD, it can infer the presence and temporal-spatial activity of the ISAC transmission, enhancing platform observability. 
		Successful CA further reveals waveform structures and system-level parameters, enabling identification of sensing signals even at low received power \cite{pace2009detecting}. 
		Such information leakage may facilitate target evasion or countermeasures, directly degrading sensing performance.
		
		These consequences differ fundamentally from classical covert or low-probability-of-detection communication, where interception mainly threatens message confidentiality or detection \cite{jiang2024physical}. 
		In sensing security aware ISAC systems, interception compromises platform observability and mission integrity even without decoding communication data, motivating the sensing-security–oriented design in this work.
	\end{remark}

	\subsection{LPI Design Guideline}\label{LPI_guideline}
	
	With the properties of intercepted signals at hand, an effective LPI design guideline should be established.
	Since the advanced ER aircraft applies both PD and CA intercept processing methods to determine the existence of the received signal, the probability of intercept, ${\rm{P}}{{\rm{r}}_{{\rm{int}}}}\left( {{{\bf{F}}_{{\rm{RF}}}},\{{\bf{F}}_k}\} \right)$, is related to the power \eqref{eq:6} and the ergodic cyclic spectrum \eqref{eq:13} of the received signal.
	To this end, we can write ${\rm{P}}{{\rm{r}}_{{\rm{int}}}}\left( {{{\bf{F}}_{{\rm{RF}}}},\{{\bf{F}}_k}\} \right)$ as a function about $P_{\rm E}$ and ${{\bf{\tilde C}}_{\rm{E}}}$ as follows.
	\begin{equation}\label{eq:14}
		{\rm{P}}{{\rm{r}}_{{\rm{int}}}}\left( {{{\bf{F}}_{{\rm{RF}}}},\{{\bf{F}}_k}\} \right) = {\cal F}\left( {{P_{\rm{E}}},{{{\bf{\tilde C}}}_{\rm{E}}}} \right),
	\end{equation}
	where the specific form of the function ${\cal F}\left( \cdot\right)$ is determined by the internal decision-making mechanisms of the advanced ER aircraft, which is unknown at the BS.
	
	Although the knowledge of the function ${\cal F}\left( \cdot\right)$ is not available, we can build some useful relationships between the probability of intercept and properties of intercepted signals.
	
	\subsubsection{Anti-PD Design Guideline}
	To combat PD, a direct and effective approach is to guarantee the received power \eqref{eq:6} as low as possible so that the preset threshold cannot be reached.
	
	\subsubsection{Anti-CA Design Guideline}
\begin{figure}[!t]
		\centering  
		\includegraphics[width=0.45\linewidth]{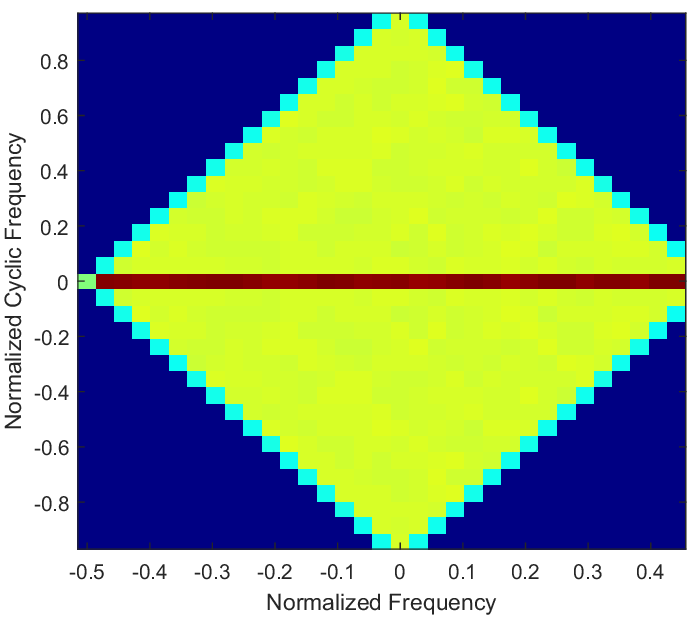}
		\vspace{-0.5em}
		\caption{Cyclic spectrum of the AWGN.}
		\label{fig:CS-AWGN}
		\vspace{-1.5em}
	\end{figure}

	To combat CA, an effective heuristic approach is to enforce the ergodic cyclic spectrum of the received signals close to that of the AWGN so that the ISAC signal can be hidden in the environmental noise.
	Due to this, the ergodic cyclic spectrum of the AWGN can be expressed as
	\begin{equation}\label{eq:15}
		{{\bf{\tilde C}}_{\rm{noise}}}\left( {m,n} \right) = {\rm{Tr}}\left\{  { {\bf{\Omega }}_0 {\bf{V}}_{{\rm{FFT}}}^H{{\bf{\Lambda }}_{m,n}}{{\bf{V}}_{{\rm{FFT}}}}} \right\},
	\end{equation}
	where ${{\bf{\Omega }}_0} = {\rm{Diag}}\left( {\sigma _{\rm{0}}^2, \ldots ,\sigma _{\rm{0}}^2} \right) \in {{\mathbb C}^{N \times N}}$ denotes the corresponding diagonal form of the general AWGN, and ${\sigma _{\rm{0}}^2}$ denotes the power spectral density.
	To better analyze the ergodic cyclic spectrum \eqref{eq:15}, we equivalently rewrite it to a more intuitive form as follows.
	\begin{equation}\label{eq:16}
		\left\{ \begin{array}{l}
			{{\bf{ \tilde C}}_{\rm{N}}}\left( {m,n} \right) = {N_0},\quad n = 0\\
			{{\bf{\tilde C}}_{\rm{N}}}\left( {m,n} \right) = \;0\;\;,\quad n \ne 0
		\end{array} \right.,
	\end{equation}
	where $N_0$ represents a constant power spectral density after processing.
	For visualization, the cyclic spectrum of the AWGN is presented in Fig. \ref{fig:CS-AWGN}.
	Based on \eqref{eq:16} and Fig.\ref{fig:CS-AWGN}, we notice that the AWGN does not exhibit any cyclostationary properties \cite{gardner1986spectral}.
	
	To enforce the ergodic cyclic spectrum of the received signal \eqref{eq:13} approaching that of the AWGN \eqref{eq:15}, we give the following insight to specify the anti-CA design objective.
	
	\begin{insight}(Anti-CA Design Objective)\label{insight:1}
		It can be observed that the only difference between the ergodic cyclic spectrum of the received signal and that of AWGN lies in the terms $\left( {\bf{R}}_\Xi + {\bf{\Omega}} \right)$ and ${\bf{\Omega}}_0$, respectively.
		Since all diagonal elements of ${\bf{\Omega}}_0$ are identical and its off-diagonal elements are zero, the ergodic cyclic spectrum of AWGN remains flat over the frequency domain.
		Accordingly, an effective strategy to render ${{\bf{\tilde C}}_{\rm E}}$ indistinguishable from ${{\bf{\tilde C}}_{\rm noise}}$ for anti-CA design is to enforce equal diagonal elements in ${\bf{R}}_\Xi$ while suppressing its off-diagonal entries.
	\end{insight}
	
	\begin{remark}{(Design Guideline Merits):}\label{remark:1}
		The state-of-the-art literature \cite{liu2022lpi,liu2023lpi,shi2025low} has also approximated the cyclic spectrum characteristic of the designed waveform to the AWGN.
		These works \cite{liu2022lpi,liu2023lpi,shi2025low} are based on optimizing certain explicit features of the cyclic spectrum, i.e., mean square error (MSE) relative to AWGN or cyclic frequency sidelobe level (CFSL), but these metrics are inherently incomplete descriptors and do not fully characterize the global structural properties of the cyclic spectrum.
		Our approach addresses this limitation by exploiting the intrinsic mathematical properties of the cyclic spectrum matrix itself.
	\end{remark}
	
	Based on \textbf{Insight \ref{insight:1}}, we further rewrite ${\bf{R}}_\Xi$ as
	\begin{equation}\label{eq:17}
		{{\bf{R}}_\Xi }\left[ {m,n} \right] =\sum\limits_{k = 1}^K {{c_k^2}{e^{\jmath 2\pi {f_k}(m - n){T_s}}}},
		m,n= 1,\dots, N,
	\end{equation}
	where ${c_k^2} = {\left| {{\beta _k}} \right|^2}\left\| {{\bf{a}}_t^H\left( {{\theta _{\rm{E}}},{f_k}} \right){{\bf{F}}_{{\rm{RF}}}}{{\bf{F}}_k}} \right\|_F^2, \forall k$.
	Now, we give the following lemma to characterize the magnitude of the ergodic cyclic spectrum ${{\bf{R}}_\Xi }$.
	\begin{lemma}\label{lemma:1}
		If $c_k^2 = c^2, \forall k$, the magnitude of ergodic cyclic spectrum \eqref{eq:17} can be expressed as
		\begin{equation}\label{eq:18}
			\left| {{{\bf{R}}_\Xi }\left[ {m,n} \right]} \right| = {c^2}\left| {\frac{{\sin \left( {\pi K\frac{{m - n}}{N}} \right)}}{{\sin \left( {\pi \frac{{m - n}}{N}} \right)}}} \right|,
		\end{equation}
		which can be regarded as nearly a diagonal matrix.
	\end{lemma}
	\begin{IEEEproof}
		Please refer to Appendix \ref{app:1}.
	\end{IEEEproof}

	\section{Sensing Security Oriented OFDM-ISAC Problem Formulation}
	
	In this section, we first derive the performance metrics, and then formulate the design problem.

	\subsection{Performance Metrics}
	
	In this work, we aim to achieve sensing security in the considered ISAC system.
	
	For communication performance, based on \eqref{eq:2}, the spectral efficiency (SE) is given by
	\begin{equation}\label{eq:19}
		{\rm{SE}}\left( {{{\bf{F}}_{{\rm{RF}}}},{{\bf{F}}_k}} \right) = \frac{1}{K}\sum\limits_{k = 1}^K {\sum\limits_{u = 1}^U {{\rm{log}_2}\left( {1 + {\rm{SIN}}{{\rm{R}}_{k,u}}} \right)} } ,
	\end{equation}
	with the SINR of the received signal
	\begin{equation}\label{eq:20}
		{\rm{SIN}}{{\rm{R}}_{k,u}}\left( {{{\bf{F}}_{{\rm{RF}}}},{{\bf{F}}_k}} \right) = \frac{{{{\left| {{\bf{h}}_{k,u}^H{{\bf{F}}_{{\rm{RF}}}}{{\bf{f}}_{k,u}}} \right|}^2}}}{{\sum\limits_{i \ne u}^U {{{\left| {{\bf{h}}_{k,u}^H{{\bf{F}}_{{\rm{RF}}}}{{\bf{f}}_{k,i}}} \right|}^2}}  + \sigma _{\rm{C}}^2}}.
	\end{equation}
	
	For radar performance, the transmit spectrum can be expressed as
	\begin{equation}\label{eq:21}
		P\left( {\theta ,{f_k}} \right) = \left\| {{\bf{a}}_t^H\left( {\theta ,{f_k}} \right){{\bf{F}}_{{\rm{RF}}}}{{\bf{F}}_k}} \right\|_F^2.
	\end{equation}
	Since the probability of detection increases monotonically with the radar SINR \cite{de2008design}, we introduce the output radar SINR as the sensing metric.
	Based on \eqref{eq:4}, the output radar SINR at the $k$-th subcarrier can be expressed as
	\begin{equation}\label{eq:22}
		\begin{aligned}
			&{\rm{SIN}}{{\rm{R}}_{\rm{r}}}\left( {{{\bf{F}}_{{\rm{RF}}}},{{\bf{F}}_k} } \right) =\\
			&{\frac{{{{\left| {{\varsigma _{{\rm{E}},k}}} \right|}^2}\left\| {{\bf{a}}_t^H\left( {{\theta _{\rm{E}}},{f_k}} \right){{\bf{F}}_{{\rm{RF}}}}{{\bf{F}}_k}} \right\|_F^2}}{{\sum\limits_{i = 1}^I\! {{{\left| {{\varsigma _{i,k}}} \right|}^2}\!\left\| {{\bf{a}}_r^H\left( {{\theta _{\rm{E}}},{f_k}} \right){{\bf{a}}_r}\left( {{\theta _i},{f_k}} \right){\bf{a}}_t^H\left( {{\theta _i},{f_k}} \right){{\bf{F}}_{{\rm{RF}}}}{{\bf{F}}_k}} \right\|_F^2} \! + \! \sigma _{\rm{R}}^2}}}.
		\end{aligned}
	\end{equation}%
	When it comes to sensing security performance, we follow the LPI design guideline in Sec. \ref{LPI_guideline}.

	\subsection{Problem Statement}
	
	In practice, the target direction is generally unknown a priori, and radar sensing is therefore performed by scanning the entire angular region using spatial beams. 
	In this scenario, the effective sensing performance depends not only on the transmit power aligned with the target direction but also on the interference caused by clutters and leakage at other scanning angles. 
	To characterize this, we define the mainlobe region as $\Theta  = \left\{ {{\theta _m}} \right\}_{m = 1}^M = \left[ {{\theta _1}, \ldots ,{\theta _M}} \right]$ and the clutters' direction region as $\Omega  = \left\{ \vartheta_s  \right\}_{s = 1}^S = \left[ {{\vartheta _1}, \ldots ,{\vartheta _S}} \right]$, with $M$ and $S$ denoting the respective number of angular sampling points.
	
	To achieve LPI against multiple threats, i.e., PD and CA, for the considered OFDM-ISAC system, we propose to design the analog beamformer, ${\bf F}_{\rm RF}$, and digital beamformer, $\{{\bf F}_k\}$, by maximizing the communication SE while meeting the sensing, power and hardware requirements.
	Mathematically, the LPI design problem can be formulated as
	\begin{subequations}
		\begin{align}
			&\mathop {\max }\limits_{{{\bf{F}}_{{\rm{RF}}}},\{ {{\bf{F}}_k}\} } {\rm{SE}}\left( {{{\bf{F}}_{{\rm{RF}}}},\{ {{\bf{F}}_k}\} } \right),\label{eq:P1-a}\\
			&\quad\;\;{\rm{s.t.}}\quad {\rm{P}}{{\rm{r}}_{{\rm{int}}}}\left( {{{\bf{F}}_{{\rm{RF}}}},\{ {{\bf{F}}_k}\} } \right) \le \varepsilon ,\label{eq:P1-b}\\
			&\qquad \qquad P\left( {{\theta _m},{f_k}} \right) \ge {\kappa _{{\rm{main}}}},\forall k,m,\label{eq:P1-c}\\
			&\qquad \qquad P\left( {{\vartheta  _s},{f_k}} \right) \le {\zeta _k},\forall k,s,\label{eq:P1-d}\\
			&\qquad\qquad\left\| {{{\bf{F}}_{{\rm{RF}}}}{{\bf{F}}_k}} \right\|_F^2 \le {{\cal P}_k},\forall k,\label{eq:P1-e}\\
			&\qquad\qquad\left| {{{\bf{F}}_{{\rm{RF}}}}\left[ {i,j} \right]} \right| = 1,\forall i,j,\label{eq:P1-f}
		\end{align}\label{eq:P1}%
	\end{subequations}
	where \eqref{eq:P1-b} denotes the LPI constraint with a low probability-of-intercept threshold $\varepsilon$, \eqref{eq:P1-c} denotes the transmit mainlobe level constraint with the threshold ${\kappa _{{\rm{main}}}}$, \eqref{eq:P1-d} denotes the constraint of nulling toward the interferences with the threshold ${\zeta _k}$, \eqref{eq:P1-e} denotes the transmit power constraint at the $k$-th subcarrier with the threshold ${{ P}_{k}}$, and \eqref{eq:P1-f} denotes the hardware constraint for the analog beamformer.
	This formulation directly follows from the spatial signal model in which interception capability is inherently direction-dependent.
	
	Note that the LPI design problem \eqref{eq:P1} is a non-convex optimization problem with an uncertain complicated constraint, constant modulus constraint, and coupling among variables.
	To tackle these difficulties, we propose an iterative approach to efficiently solve this problem in the following sections.
	
	\section{LPI Design Against Multi-Intercept Threats}
	
	In this section, we first reformulate the original problem into a more tractable form, and then propose an effective solver to jointly design the radar filter and hybrid transmitter.

	\subsection{Problem Transformation}
	
	\subsubsection{Simplification of Objective function \eqref{eq:P1-a}}
	
	To handle the complex objective function \eqref{eq:P1-a}, we adopt the weighted minimum mean square error (WMMSE)-Rate relationship \cite{xu2024enhancing}
	\begin{equation}\label{eq:24}
		{{\rm{log}_2}\left( {1 + {\rm{SIN}}{{\rm{R}}_{k,u}}} \right)} = {\rm{lo}}{{\rm{g}}_2}{\omega _{k,u}} - {\omega _{k,u}}{{\rm{E}}_{u,k}}\left( {{{\bf{F}}_{{\rm{RF}}}},{{\bf{F}}_k}} \right) + 1,
	\end{equation}
	where ${{\rm E}_{u,k}}\left( {{{\bf{F}}_{{\rm{RF}}}},{{\bf{F}}_k}} \right) = {\mathbb E}\{ {{{\left| {{s_{k,u}} - {\kappa _{k,u}}{r_{k,u}}} \right|}^2}} \} = {\left| {{\kappa _{k,u}}} \right|^2}( {\| {{\bf{h}}_{k,u}^H{{\bf{F}}_{{\rm{RF}}}}{{\bf{F}}_k}} \|_F^2 + \sigma _{\rm{C}}^2} ) - 2\Re \{ {{\kappa _{k,u}}{\bf{h}}_{k,u}^H{{\bf{F}}_{{\rm{RF}}}}{{\bf{f}}_{k,u}}} \} + 1$ is the MSE, and the equalizer $\kappa_{k,u}$ and the weight $\omega_{k,u}$ are derived by
	\begin{subequations}
		\begin{align}
			&{\kappa _{k,u}} = \frac{{{\bf{f}}_{k,u}^H{\bf{F}}_{{\rm{RF}}}^H{{\bf{h}}_{k,u}}}}{{\left\| {{\bf{h}}_{k,u}^H{{\bf{F}}_{{\rm{RF}}}}{{\bf{F}}_k}} \right\|_F^2 + \sigma _{\rm{C}}^2}}, \forall k,u,\\
			&{\omega _{k,u}} = \frac{1}{{{{\rm{E}}_{u,k}}\left( {{{\bf{F}}_{{\rm{RF}}}},{{\bf{F}}_k}} \right)}}, \forall k,u.
		\end{align}\label{eq:25}
	\end{subequations}
	
	\subsubsection{Specification of Constraint \eqref{eq:P1-b}}
	
	It is noticed that the constraint \eqref{eq:P1-b} involves a function ${\cal F}\left( \cdot\right)$.
	In the practical scenario, the specific information of the function ${\cal F}\left( \cdot\right)$ tends to be unknown by the OFDM-ISAC BS, which makes the original problem extremely tricky.
	To address this issue, we aim to transform the constraint \eqref{eq:P1-b} into a more tractable form.
	
	Based on the anti-PD design guideline in Sec. \ref{LPI_guideline}, the power of the received signals should be suppressed while the constraint of mainlobe level \eqref{eq:P1-c} should be guaranteed.
	Based on the anti-CA design guideline in \textbf{Lemma \ref{lemma:1}}, the diagonal elements of $\bf \Xi$ should be identical, leading to the following problem of LPI design against PD and CA:
	\begin{equation}\label{eq:26}
		{\varpi _k}\left\| {{\bf{a}}_t^H\left( {{\theta _m},{f_k}} \right){{\bf{F}}_{{\rm{RF}}}}{{\bf{F}}_k}} \right\|_F^2 = \eta ,\forall k,m,
	\end{equation}
	where $\eta$ represents the weighted mainlobe level, and ${\varpi _k}$ is a weight proportional to $\beta_k$ with the reference to the center frequency.
	
	Now, the original LPI design problem \eqref{eq:P1} can be reformulated as
	\begin{subequations}
		\begin{align}
			&\mathop {\min }\limits_{{{\bf{F}}_{{\rm{RF}}}},\{ {{\bf{F}}_k}\} } \sum\limits_{k = 1}^K {\sum\limits_{u = 1}^U {{{\rm{E}}_{u,k}}\left( {{{\bf{F}}_{{\rm{RF}}}},{{\bf{F}}_k}} \right)} } ,\label{eq:P2-a}\\
			&\quad\;\;{\rm{s}}.{\rm{t}}.\quad {\varpi _k}\left\| {{\bf{a}}_t^H\left( {{\theta _m},{f_k}} \right){{\bf{F}}_{{\rm{RF}}}}{{\bf{F}}_k}} \right\|_F^2 = \eta ,\forall k,m,\label{eq:P2-b}\\
			&\qquad\qquad \eqref{eq:P1-d}-\eqref{eq:P1-f}.
		\end{align}\label{eq:P2}%
	\end{subequations}
	
	\subsubsection{Problem Reformulation}
	
	To decouple the hybrid digital and analog beamformers in the objective function \eqref{eq:P2-a} and constraints \eqref{eq:26}, \eqref{eq:P1-d}-\eqref{eq:P1-f}, we introduce several auxiliary variables $\{{{{\bf{Y}}_k}\} ,\{ {{\bf{V}}_{k,m}}\} ,\{ {{\bf{G}}_{k,s}}\} ,\{ {{\bf{T}}_{k,u}}\} }$ to rewrite the problem \eqref{eq:P2} as
	\begin{subequations}
		\begin{align}
			&\mathop {\min }\limits_{{\scriptstyle {{\bf{F}}_{{\rm{RF}}}},\{ {{\bf{F}}_k}\} ,\{{{{\bf{Y}}_k}\} ,\hfill}\atop
					{\scriptstyle\{ {{\bf{V}}_{k,m}}\} ,\{ {{\bf{G}}_{k,s}}\} ,\{ {{\bf{T}}_{k,u}}}\} \hfill}} 
			\;\sum\limits_{k = 1}^K 
			{\sum\limits_{u = 1}^U {{{\rm{E}}_{u,k}}\left( {{{\bf{T}}_{k,u}}} \right)} },\label{eq:P3-a}\\
			&\qquad\qquad{\rm{s.t.}}\; {\varpi _k}\left\| {{\bf{a}}_t^H\left( {{\theta _m},{f_k}} \right){{\bf{V}}_{k,m}}} \right\|_F^2 = \eta ,\forall k,m,\label{eq:P3-b}\\
			&\qquad \qquad\quad\left\| {{\bf{a}}_t^H\left( {{\vartheta _s},{f_k}} \right){{\bf{G}}_{k,s}}} \right\|_F^2\; \le {\zeta _k},\forall k,s,\label{eq:P3-c}\\
			&\qquad \qquad\quad \left\| {{{\bf{Y}}_k}} \right\|_F^2 \le {{\cal P}_k},\forall k,\label{eq:P3-d}\\
			&\qquad \qquad\quad\left| {{{\bf{F}}_{{\rm{RF}}}}\left[ {i,j} \right]} \right| = 1,\forall i,j,\label{eq:P3-e}\\
			&\qquad \qquad\quad{{\bf{Y}}_k} = {{\bf{V}}_{k,m}} = {{\bf{G}}_{k,s}} = {{\bf{T}}_{k,u}} = \;{{\bf{F}}_{{\rm{RF}}}}{{\bf{F}}_k},\label{eq:P3-g}
		\end{align}\label{eq:P3}%
	\end{subequations}
	whose associated augmented Lagrangian (AL) function via penalizing the equality constraints \eqref{eq:P3-g} is given in \eqref{eq:29}, where $\{ {{{\bf{D}}_{1,k}}}\}$,$ \{ {{{\bf{D}}_{2,k,m}}}\}$, $\{ {{{\bf{D}}_{3,k,s}}}\}$,$\{{{{\bf{D}}_{4,k,u}}}\}$ and $\rho_1,~\rho_2,~\rho_3,~\rho_4>0$ are dual variables and corresponding penalty parameters, respectively.
	\begin{figure*}[ht!]
		\centering
		\begin{equation}\label{eq:29}
			\begin{aligned}
				&{\cal L}\left( {\{ {{\bf{Y}}_k}\} ,\{ {{\bf{V}}_{k,m}}\} ,\{ {{\bf{G}}_{k,s}}\} ,\{ {{\bf{T}}_{k,u}}\} ,{{\bf{F}}_{{\rm{RF}}}},\{ {{\bf{F}}_k}\} ,\{ {{\bf{D}}_{n,k,l}}\} } \right)\\
				&\qquad = \sum\limits_{k = 1}^K {\sum\limits_{u = 1}^U {{{\rm{E}}_{u,k}}\left( {{{\bf{T}}_{k,u}}} \right)} }  + \frac{{{\rho _1}}}{2}\sum\limits_{k = 1}^K {\left\| {{{\bf{Y}}_k} - {{\bf{F}}_{{\rm{RF}}}}{{\bf{F}}_k} + {{\bf{D}}_{1,k}}} \right\|_F^2}  + \frac{{{\rho _2}}}{2}\sum\limits_{k = 1}^K {\sum\limits_{m = 1}^M {\left\| {{{\bf{V}}_{k,m}} - {{\bf{Y}}_k} + {{\bf{D}}_{2,k,m}}} \right\|_F^2} }  \\
				&\qquad\quad + \frac{{{\rho _3}}}{2}\sum\limits_{k = 1}^K {\sum\limits_{s = 1}^S {\left\| {{{\bf{G}}_{k,s}} - {{\bf{Y}}_k} + {{\bf{D}}_{3,k,s}}} \right\|_F^2} }  + \frac{{{\rho _4}}}{2}\sum\limits_{k = 1}^K {\sum\limits_{u = 1}^U {\left\| {{{\bf{T}}_{k,u}} - {{\bf{Y}}_k} + {{\bf{D}}_{4,k,u}}} \right\|_F^2} } .
			\end{aligned}
		\end{equation}
		\hrule
		\vspace{-1em}
	\end{figure*}
	
	\vspace{-1em}
	\subsection{Alternating Optimization (AO) of Problem \eqref{eq:P3}}
	Now, the problem \eqref{eq:P3} can be iteratively solved as follows.
	
	\subsubsection{Update of ${\{ {{\bf{Y}}_k}\} }$}
	With other variables fixed, ${\{ {{\bf{Y}}_k}\} }$ can be updated by solving
	\begin{equation}
		\begin{aligned}
			&\mathop {\min }\limits_{\{ {{\bf{Y}}_k}\} } \;\frac{{{\rho _1}}}{2}\sum\limits_{k = 1}^K {\left\| {{{\bf{Y}}_k} - {{\bf{F}}_{{\rm{RF}}}}{{\bf{F}}_k} + {{\bf{D}}_{1,k}}} \right\|_F^2} \\
			&\qquad\;\;+ \frac{{{\rho _2}}}{2}\sum\limits_{k = 1}^K {\sum\limits_{m = 1}^M {\left\| {{{\bf{V}}_{k,m}} - {{\bf{Y}}_k} + {{\bf{D}}_{2,k,m}}} \right\|_F^2} } \\
			&\qquad\;\;+ \frac{{{\rho _3}}}{2}\sum\limits_{k = 1}^K {\sum\limits_{s = 1}^S {\left\| {{{\bf{G}}_{k,s}} - {{\bf{Y}}_k} + {{\bf{D}}_{3,k,s}}} \right\|_F^2} }  \\
			&\qquad\;\; + \frac{{{\rho _4}}}{2}\sum\limits_{k = 1}^K {\sum\limits_{u = 1}^U {\left\| {{{\bf{T}}_{k,u}} - {{\bf{Y}}_k} + {{\bf{D}}_{4,k,u}}} \right\|_F^2} },\\
			&\;{\rm{s.t.}}\;\;\left\| {{{\bf{Y}}_k}} \right\|_F^2 \le {{\cal P}_k},\forall k,
		\end{aligned}\label{eq:30}
	\end{equation}
	which is also a convex quadratically constrained quadratic programming (QCQP) whose optimal solutions can be obtained by analyzing the KKT conditions.
	Then, the closed-form solution to the problem \eqref{eq:30} is derived by
	\begin{equation}\label{eq:31}
		{{\bf{Y}}_k}\left( {{\chi _k}} \right) = {\left( {{\bf{C}} + 2{\chi _k}{{\bf{I}}_{{M_t}}}} \right)^{ - 1}}{{\bf{B}}_k}, \forall k,
	\end{equation}
	where ${\bf{C}} = \left( {{\rho _1} + M{\rho _2} + S{\rho _3} + U{\rho _4}} \right){{\bf{I}}_{{M_t}}}$, ${{\bf{B}}_k} = {\rho _1}\left( {{{\bf{F}}_{{\rm{RF}}}}{{\bf{F}}_k} - {{\bf{D}}_{1,k}}} \right) + {\rho _2}\sum\nolimits_{m = 1}^M {\left( {{{\bf{V}}_{k,m}} + {{\bf{D}}_{2,k,m}}} \right)}  + {\rho _3}\sum\nolimits_{s = 1}^S {\left( {{{\bf{G}}_{k,s}} + {{\bf{D}}_{3,k,s}}} \right)}  + {\rho _4}\sum\nolimits_{u = 1}^U {\left( {{{\bf{T}}_{k,u}} + {{\bf{D}}_{4,k,u}}} \right)} $. 
	The multiplier ${{\chi _k}}$ can be solved by substituting \eqref{eq:31} into the power constraint using the bisection method.

	\subsubsection{Update of ${\{ {{\bf{V}}_{k,m}}\} }$}
	With other variables fixed, ${\{ {{\bf{V}}_{k,m}}\} }$ can be updated by solving
	\begin{equation}
		\begin{aligned}
			&\mathop {\min }\limits_{\{ {{\bf{V}}_{k,m}}\} } \sum\limits_{k = 1}^K {\sum\limits_{m = 1}^M {\left\| {{{\bf{V}}_{k,m}} - {{\bf{Y}}_k} + {{\bf{D}}_{2,k,m}}} \right\|_F^2} } ,\\
			&\quad{\rm{s.t.}}\;\; \left\| {{\bf{a}}_t^H\left( {{\theta _m},{f_k}} \right){{\bf{V}}_{k,m}}} \right\|_F^2 = \frac{\eta }{{{\varpi _k}}},\forall k,m.
		\end{aligned}\label{eq:32}
	\end{equation}
	Note that this problem is also a QCQP whose closed-form solution can be derived by
	\begin{equation}\label{eq:33}
		{{\bf{V}}_{k,m}} = \left( {{{\bf{I}}_{{M_t}}} - \frac{{{\mu _{k,m}}{{\bf{A}}_{m,k}}}}{{1 + {\mu _{k,m}}}}} \right)\left( {{{\bf{Y}}_k} - {{\bf{D}}_{2,k,m}}} \right), \forall k,m,
	\end{equation}
	where ${\mu _{k,m}} = \sqrt {\frac{{{\varpi _k}}}{\eta }} {\left\| {{\bf{a}}_t^H\left( {{\theta _m},{f_k}} \right)\left( {{{\bf{Y}}_k} - {{\bf{D}}_{2,k,m}}} \right)} \right\|_F} - 1$, with defining ${{\bf{A}}_{m,k}} = {{\bf{a}}_t}\left( {{\theta _m},{f_k}} \right){\bf{a}}_t^H\left( {{\theta _m},{f_k}} \right),\forall m,k$.
	
	\subsubsection{Update of ${\{ {{\bf{G}}_{k,s}}\} }$}
	With other variables fixed, ${\{ {{\bf{G}}_{k,s}}\} }$ can be updated by solving
	\begin{equation}\label{eq:34}
		\begin{aligned}
			&\mathop {\min }\limits_{\{ {{\bf{G}}_{k,s}}\} } \sum\limits_{k = 1}^K {\sum\limits_{s = 1}^S {\left\| {{{\bf{G}}_{k,s}} - {{\bf{Y}}_k} + {{\bf{D}}_{3,k,s}}} \right\|_F^2} },\\
			&\;\;\;{\rm{s.t.}}\;\left\| {{\bf{a}}_t^H\left( {{\vartheta _s},{f_k}} \right){{\bf{G}}_{k,s}}} \right\|_F^2\; \le {\zeta _k},\forall k,s,
		\end{aligned}
	\end{equation}
	which is a QCQP with the closed-form solution:
	\begin{equation}\label{eq:35}
		{{\bf{G}}_{k,s}}\left( {{\lambda _{3,k,s}}} \right) = {\left( {{{\bf{I}}_{{M_t}}} + {\lambda _{3,k,s}}{{\bf{\Lambda }}_{s,k}}} \right)^{ - 1}}\left( {{{\bf{Y}}_k} - {{\bf{D}}_{3,k,s}}} \right),
	\end{equation}
	where ${{\bf{\Lambda }}_{s,k}} = {{\bf{a}}_t}\left( {{\vartheta _s},{f_k}} \right){\bf{a}}_t^H\left( {{\vartheta _s},{f_k}} \right), \forall s,k$, and ${{\lambda _{3,k,s}}}$ is the corresponding multiplier.
	Based on this, we consider two cases:
	\begin{itemize}
		\item Case 1: ${\lambda _{3,k,s}} = 0$, the solution to the subproblem \eqref{eq:35} is 
		${{\bf{G}}_{k,s}} = {{\bf{Y}}_k} - {{\bf{D}}_{3,k,s}}$,
		which satisfies $\left\| {{\bf{a}}_t^H\left( {{\vartheta _s},{f_k}} \right){{\bf{G}}_{k,s}}}\right\|_F^2 \le {\zeta _k}$.
		\item Case 2: ${\lambda _{3,k,s}} > 0$, the solution to the subproblem \eqref{eq:35} should satisfy the following equality relationship:
		\begin{equation}\label{eq:36}
			\left\| {{\bf{a}}_t^H\left( {{\vartheta _s},{f_k}} \right){{\bf{G}}_{k,s}}\left( {{\lambda _{3,k,s}}} \right)} \right\|_F^2\; = {\zeta _k},
		\end{equation}
		where the optimal $\lambda _{3,k,s}^ \star $ can be found using bisection method. 
		After substituting $\lambda _{3,k,s}^ \star $ into \eqref{eq:35}, the solution to ${{\bf{G}}_{k,s}}$ is obtained.
	\end{itemize}
	
	\subsubsection{Update of ${\{ {{\bf{T}}_{k,u}}\} }$}
	With other variables fixed, ${\{ {{\bf{T}}_{k,u}}\} }$ can be updated by solving
	\begin{equation}\label{eq:37}
		\mathop {\min }\limits_{\{ {{\bf{T}}_{k,u}}\} } \!\sum\limits_{k = 1}^K\! {\sum\limits_{u = 1}^U \!{{{\rm{E}}_{u,k}}\left( {{{\bf{T}}_{k,u}}} \right)} } \! + \!\frac{{{\rho _4}}}{2}\sum\limits_{k = 1}^K \!{\sum\limits_{u = 1}^U \!{\left\| {{{\bf{T}}_{k,u}} \!\!-\!\! {{\bf{Y}}_k} \!+ \!{{\bf{D}}_{4,k,u}}} \right\|_F^2} },
	\end{equation}
	whose optimal solutions can be obtained by applying the KKT conditions:
	\begin{equation}\label{eq:38}
		\begin{aligned}
			{\bf{T}}_{k,u}^ \star  =& {\left( {2{{\left| {{\kappa _{k,u}}} \right|}^2}{{\bf{h}}_{k,u}}{\bf{h}}_{k,u}^H + {\rho _4}{{\bf{I}}_{{M_t}}}} \right)^{ - 1}} \\
			&\cdot \left( {2\kappa _{k,u}^*{{\bf{h}}_{k,u}}{\bf{e}}_u^T + {\rho _4}\left( {{{\bf{Y}}_k} - {{\bf{D}}_{4,k,u}}} \right)} \right),
		\end{aligned}
	\end{equation}
	where ${{\bf{e}}_u} = {\left[ {0, \ldots ,0,{1_{(u)}},0, \ldots ,0} \right]^T}$.

	\subsubsection{Update of ${\{ {{{\bf{F}}_k}}\} }$}
	
	With other variables fixed, ${\bf{F}}_k$ can be updated by solving an unconstrained subproblem
	\begin{equation}\label{eq:39}
		\mathop {\min }\limits_{{{\bf{F}}_k}} \;\left\| {{\bf{Z}}_k} - {{\bf{F}}_{{\rm{RF}}}}{{\bf{F}}_k} \right\|_F^2,
	\end{equation}
	where ${{\bf{Z}}_k} = {{\bf{Y}}_k} + {{\bf{D}}_{1,k}},\forall k$.
	Taking the first-order optimality condition, its solution can be directly derived by 
	\begin{equation}\label{eq:40}
		{{\bf{F}}_k} = {\left( {{\bf{F}}_{{\rm{RF}}}^H{{\bf{F}}_{{\rm{RF}}}}} \right)^{ - 1}}{\bf{F}}_{{\rm{RF}}}^H{\bf{Z}}_k.
	\end{equation}

	\subsubsection{Update of ${{{\bf{F}}_{{\rm{RF}}}}}$}

	With other variables fixed, ${{{\bf{F}}_{{\rm{RF}}}}}$ can be updated by solving
	\begin{equation}\label{eq:41}
		\mathop {\min }\limits_{{{\bf{F}}_{{\rm{RF}}}}} \;\;\sum\limits_{k = 1}^K {\left\| {{{\bf{Z}}_k} - {{\bf{F}}_{{\rm{RF}}}}{{\bf{F}}_k}} \right\|_F^2} ,
		\;{\rm{s.t.}}\; \left| {{{\bf{F}}_{{\rm{RF}}}}\left[ {i,j} \right]} \right| = 1,\forall i,j.
	\end{equation}
	This subproblem is hard to solve due to the discrete constraint involved in the constant modulus hardware requirement.
	To tackle this difficulty, we rewrite the constant modulus constraint into an equivalent form:
	\begin{equation}\label{eq:42}
		{{\bf{F}}_{{\rm{RF}}}}\left[ {i,j} \right] = \exp \left( { - \jmath {\vartheta _{i,j}}} \right),{\vartheta _{i,j}} \in \left( {0,2\pi } \right].
	\end{equation}
	The constraint \eqref{eq:42} indicates that the elements of ${\bf F}_{\rm RF}$ lie on a closed unit circle, which is a closed set.
	Therefore, the subproblem \eqref{eq:41} can be solved under a cyclic coordinate descent (CCD) framework \cite{xu2024task}.
	
	\subsubsection{Update of Dual Variables}
	
	With other variables fixed, the dual variables can be updated by
	\begin{subequations}
		\begin{align}
			&{\bf{D}}_{1,k}^{l + 1} = {\bf{Y}}_k^{l + 1} - {\bf{F}}_{{\rm{RF}}}^{l + 1}{\bf{F}}_k^{l + 1} + {\bf{D}}_{1,k}^l,\\
			&{\bf{D}}_{2,k,m}^{l + 1} = {\bf{V}}_{k,m}^{l + 1} - {\bf{Y}}_k^{l + 1} + {\bf{D}}_{2,k,m}^l,\\
			&{\bf{D}}_{3,k,s}^{l + 1} = {\bf{G}}_{k,s}^{l + 1} - {\bf{Y}}_k^{l + 1} + {\bf{D}}_{3,k,s}^l,\\
			&{\bf{D}}_{4,k,u}^{l + 1} = {\bf{T}}_{k,u}^{l + 1} - {\bf{Y}}_k^{l + 1} + {\bf{D}}_{4,k,u}^l.
		\end{align}\label{eq:43}
	\end{subequations}%
	
	Moreover, the residual errors can be defined as
	\begin{subequations}
		\begin{align}
			&{\text{Residual Error 1}} = \frac{1}{K}\sum\limits_{k = 1}^K {\left\| {{{\bf{Y}}_k} - {{\bf{F}}_{{\rm{RF}}}}{{\bf{F}}_k}} \right\|_F^2},\\
			&{\text{Residual Error 2}} = \frac{1}{M}\frac{1}{K}\sum\limits_{k = 1}^K {\sum\limits_{m = 1}^M {\left\| {{{\bf{V}}_{k,m}} - {{\bf{Y}}_k}} \right\|_F^2} },\\
			&{\text{Residual Error 3}} = \frac{1}{S}\frac{1}{K}\sum\limits_{k = 1}^K {\sum\limits_{s = 1}^S {\left\| {{{\bf{G}}_{k,s}} - {{\bf{Y}}_k}} \right\|_F^2} },\\
			&{\text{Residual Error 4}} = \frac{1}{U}\frac{1}{K}\sum\limits_{k = 1}^K {\sum\limits_{u = 1}^U {\left\| {{{\bf{T}}_{k,u}} - {{\bf{Y}}_k}} \right\|_F^2} }.
		\end{align}\label{eq:44}%
	\end{subequations}

	\subsection{Summary}

	\setlength{\textfloatsep}{0.5em}
	\begin{algorithm}[t]
		\caption{Sensing Security Oriented ISAC Design}
		\label{alg:1}
		\KwIn{System parameters, $\{{\bf h}_{k,u}\}$.}
		\KwOut{${\bf F}_{\rm RF}$, $\{{\bf F}_k\}$.}  
		\BlankLine
		\textbf{Initialize:} ${\bf F}_{\rm RF}^{0}$, $\{{\bf F}_k^{0}\}$, $l = 0$;\\ 
		\While{No Convergence}{
			$l = l + 1$;\\
			Update $\{{\bf{Y}}_k^{l + 1}\}$ by \eqref{eq:31}.\\
			Update $\{{\bf{V}}_{k,m}^{l + 1}\}$ by solving \eqref{eq:33}.\\
			Update $\{{\bf{G}}_{k,s}^{l + 1}\}$ by \eqref{eq:35}.\\
			Update $\{ {\bf{T}}_{k,u}^{l + 1}\}$ by solving \eqref{eq:38}.\\
			Update $\{{\bf{F}}_k^{l + 1}\}$ by \eqref{eq:40}.\\
			Update ${\bf{F}}_{{\rm{RF}}}^{l + 1}$ by \eqref{eq:42} under CCD framework.\\
			Update dual variables by \eqref{eq:43}.
		}
		$\{{\bf F}_k\} = \{{\bf F}_k^{l}\}$ and ${\bf F}_{\rm RF} = {\bf F}_{\rm RF}^{l}$.
	\end{algorithm}

	According to the derivations above, we summarize the proposed algorithm for the anti-PD\&CA LPI design for secure OFDM-ISAC systems in \textbf{Algorithm \ref{alg:1}}.
	
	Then, we analyze the computational complexity of the proposed algorithm.
	Updating $\{{\bf Y}_k\}$ needs a complexity of ${\cal O}\left( {KM_t^3} \right)$, updating ${\{ {{\bf{V}}_{k,m}}\} }$ needs a complexity of ${\cal O}\left( {MKM_t^3} \right)$, updating ${\{ {{\bf{G}}_{k,s}}\} }$ needs a complexity of ${\cal O}\left( {SKM_t^3} \right)$, and updating ${\{ {{\bf{T}}_{k,u}}\} }$ needs a complexity of ${\cal O}\left( {UKM_t^3} \right)$.
	Besides, updating ${\bf{F}}_k$ needs a complexity of ${\cal O}\left( {N_{\rm RF}^3} \right)$, and updating ${{{\bf{F}}_{{\rm{RF}}}}}$ needs a complexity of ${\cal O}\left( {{N_{{\rm{CCD}}}}K{M_t}N_{\rm RF}^2U} \right)$, with $N_{\rm CCD}$ denoting the number of CCD iterations.
	To sum up, the overall complexity of \textbf{Algorithm \ref{alg:1}} is ${\cal O}\left( {{N_{{\rm{AO}}}}\left( {{N_{{\rm{CCD}}}}K{M_t}N_{\rm RF}^2U + MKM_t^3} \right)} \right)$, with ${N_{{\rm{AO}}}}$ denoting the number of AO iterations.

	\section{Numerical Simulations}
	
	In this section, we provide numerical simulations to evaluate the performance of the proposed LPI design for the secure OFDM-ISAC system against multiple threats.

	\subsection{Simulation Setup}
	
	\subsubsection{Parameter Setting}
	Unless otherwise specified, we assume that the BS is equipped with $M_t = M_r = 32$ transmit/receive antennas, where $N_{\rm RF} = 4$ RF chains are used.
	The transmit power is set as ${\cal P}_k = {\cal P} = 30\text{dBm},\forall k$.
	There are $U=4$ communication users and the communication noise power is $\sigma_{\rm C}^2 = -70$dBm.
	Besides, the CSI matrix from the OFDM-ISAC BS to the downlink communication user is constructed by the mmWave channel model in \cite{10858124,cheng2021hybrid,li2020dynamic}.
	The communication symbols are extracted from binary phase shift keying (BPSK) modulation.
	We consider $K=32$ subcarriers in this OFDM-ISAC system, where the system bandwidth is $B=2.56$GHz and the center frequency is $f_c = 24$GHz.
	In this case, the percent bandwidth $B\% = \frac{B}{f_c} = 10.67\%$, which indicates a typical wideband OFDM setting.
	The length of OFDM symbol is set as $N_{\rm symbol} = 32$ and the length of CP is set as $N_{\rm cp} = 0.125 N_{\rm symbol}$.
	Moreover, the sensing noises for BS and ER aircraft are assumed as $\sigma_{\rm R}^2 = \sigma_{\rm E}^2 = -70$dBm.
	Besides, the variance of the complex channel coefficient from BS to ER aircraft at the center frequency is assumed as $-100$dBm.
	The sampling rate is assumed $f_s = 2.56$GHz, the number of sampling points is $N = 34$ and the length of the analysis window is $W = 2$.
	The sampling number of the angular region $\left[-90^\circ, 90^\circ\right]$ is $D = 181$.
	Assume that the target is located at $26^\circ$ and the clutters are located at $-30^\circ$ and $60^\circ$, with the variance of complex amplitudes for ER aircraft and interference $\sigma_{{\rm E},k}^2 = -70$dBm and $\sigma_{i,k}^2 = -50$dBm, respectively. 
	The mainlobe region is set as $\Theta = \left[21^\circ, 29^\circ\right]$, the clutters' direction region is set as $\Omega = \left[ { - {{32}^\circ }, - {{28}^\circ }} \right] \cup \left[ {{{58}^\circ },{{62}^\circ }} \right]$.
	The weighted mainlobe level threshold is set as $\eta = 20$dBm and the nulling level threshold is set as ${\zeta _k} = \zeta = 0\text{dBm}, \forall k$.
	
	\subsubsection{Benchmark Specification}
	For comparison purposes, several benchmarks are included in this paper.
	\begin{itemize}
		\item \textbf{HBF-Comm.Only:} This scheme considers a BS with HBF that only works for downlink communication.
		\item \textbf{HBF-RadarOnly:} This scheme considers a BS with HBF that only works for target sensing.
		\item \textbf{FD-ISAC:} This scheme considers a secure BS with FD beamforming structure for ISAC against both PD and CA intercept threats, as the upper bound of the proposed HBF-ISAC scheme.
		\item \textbf{TS-Based HBF-ISAC:} This scheme considers a secure BS with HBF structure for ISAC against PD and CA intercept threats, where the HBF is designed based on the two-stage (TS) method \cite{yu2016alternating}.
	\end{itemize}

	\subsection{Simulation Results}

	\subsubsection{Convergence of the Proposed Algorithm}
	
	\begin{figure}[!t]
		\centering
		\subfigure[]{
			\includegraphics[width=0.8\linewidth]{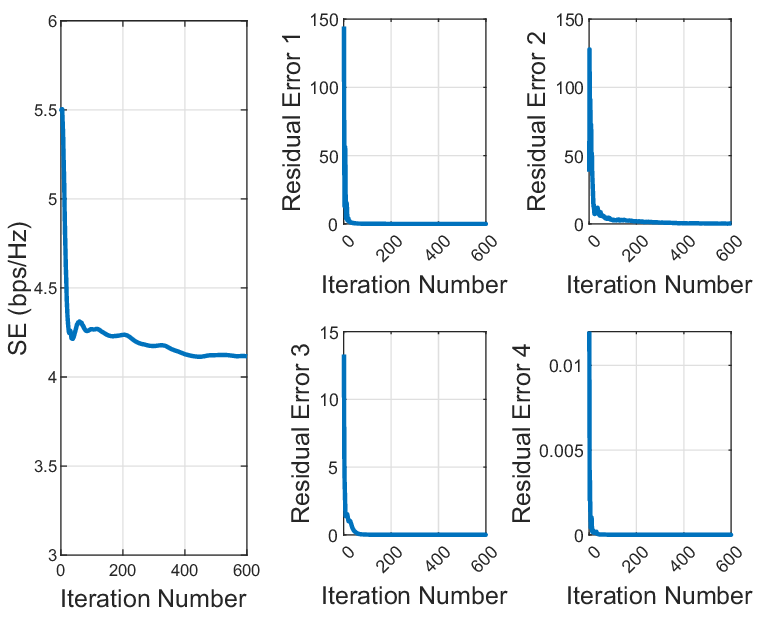} 
			\label{fig:3-1}	}
		\subfigure[]{
			\includegraphics[width=0.8\linewidth]{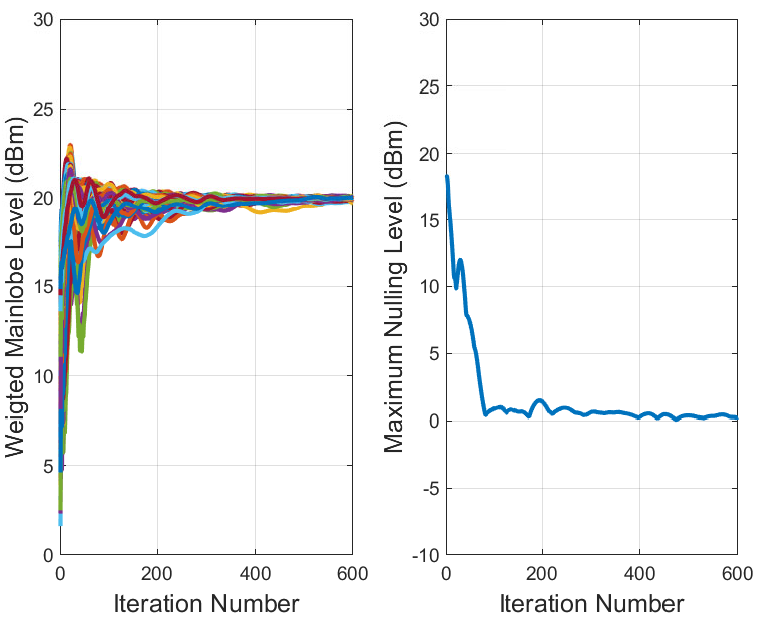} 
			\label{fig:3-2}	}
		\vspace{-0.5em}
		\caption{Convergence performance of the proposed algorithm: (a) Objective function/residual errors versus the iteration number. (b) Weighted mainlobe level versus the iteration number; maximum nulling level versus the iteration number.}
		\label{fig:3}
	\end{figure}
	
	Fig. \ref{fig:3} illustrates the convergence behavior of the proposed algorithm.
	As shown in Fig. \ref{fig:3-1}, the SE decreases and eventually converges to a stable value, while the residual error curves drop rapidly and approach zero.
	Furthermore, Fig. \ref{fig:3-2} shows that the weighted mainlobe level is first facilitated and then stabilized at the preset threshold $\eta = 20$ dBm.
	Meanwhile, the maximum nulling level initially decreases and finally remains constant at the threshold $\zeta = 0$ dBm.
	These results demonstrate the convergence of the proposed HBF-ISAC algorithm.
	
	\subsubsection{Impact of Weighted Mainlobe Level and Nulling Level}
	
	\begin{figure}[!t]
		\centering  
		\includegraphics[width=0.75\linewidth]{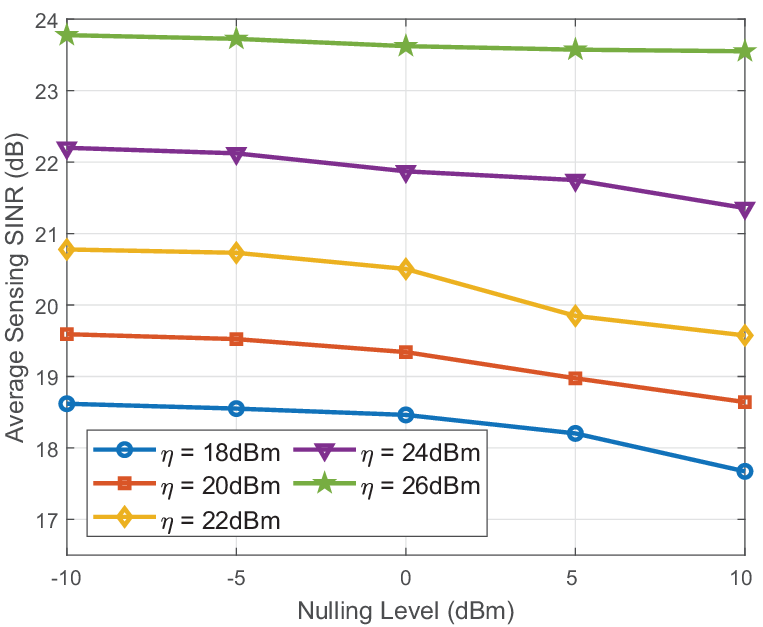}
		\vspace{-0.5em}
		\caption{Sensing SINR versus the nulling level, with different weighted mainlobe level $\eta$.}
		\label{fig:4}
	\end{figure}
	
	Fig. \ref{fig:4} plots the sensing SINR versus the nulling level with different weighted mainlobe level $\eta = 18, 20, 22, 24, 26$dBm.
	We can observe that the sensing SINRs are decreasing as the nulling level increases, and the design with a higher weighted mainlobe level leads to a higher sensing SINR.
	This demonstrates that the proposed algorithm can effectively guarantee the sensing SINR by controlling the weighted mainlobe level and nulling level.
	
	\subsubsection{Performance Comparison of Different Schemes}
	
	\begin{figure}[!t]
		\centering  
		\includegraphics[width=0.75\linewidth]{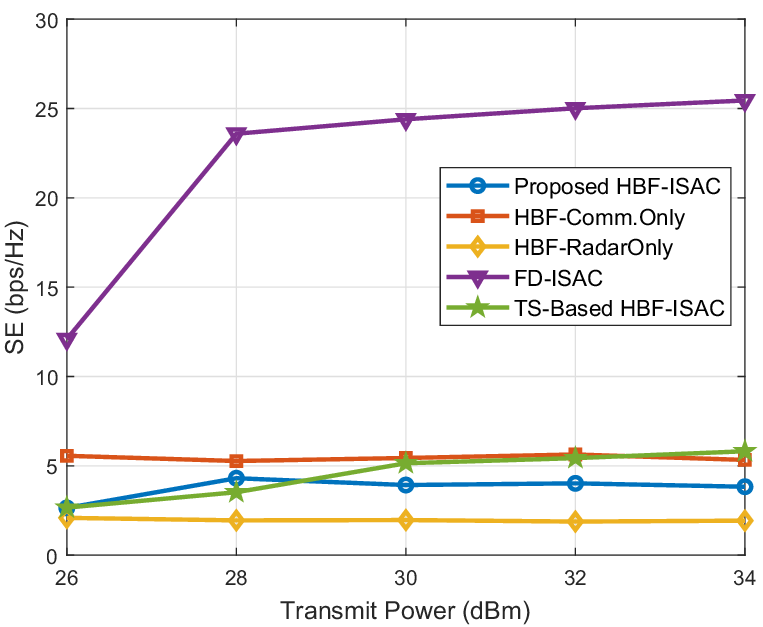}
		\vspace{-0.5em}
		\caption{Comparison of different schemes in terms of the SE versus transmit power.}
		\label{fig:5}
	\end{figure}
	
	Then, we compare the performance of different design schemes.
	Specifically, Fig. \ref{fig:5} plots the curves of the communication SE versus the transmit power.
	It can be seen that the SE achieved by the FD-ISAC design is the highest, followed by the HBF-Comm.Only design in the second position at around $5$bps/Hz.
	The SE realized by the proposed HBF-ISAC design increases as the transmit power is increased from $26$dBm to $28$dBm, and then maintains the SE at around $4$bps/Hz.
	This shows that the proposed HBF-ISAC design can realize a satisfactory communication performance closer to that of the HBF-Comm.Only design.
	
	\begin{figure}[!t]
		\centering  
		\includegraphics[width=0.75\linewidth]{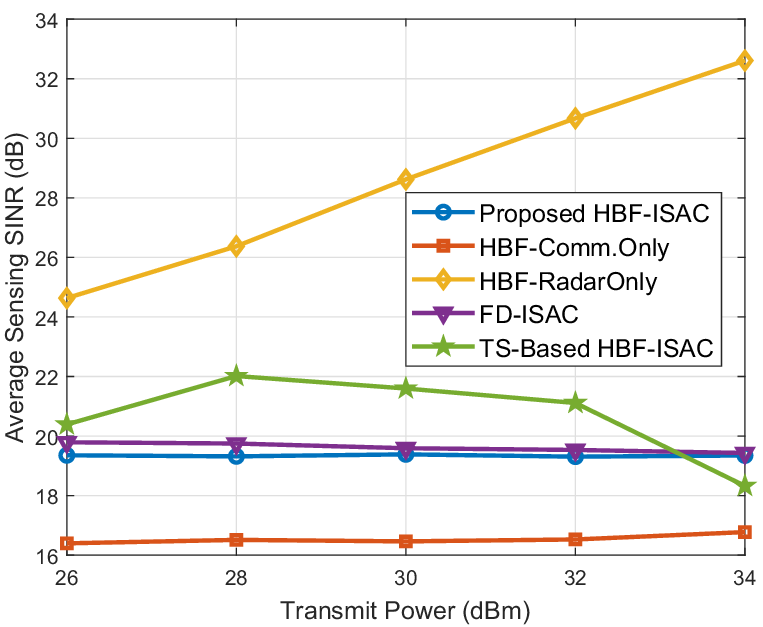}
		\vspace{-0.5em}
		\caption{Comparison of different schemes in terms of the sensing SINR versus transmit power.}
		\label{fig:6}
	\end{figure}
	
	Fig. \ref{fig:6} plots the curves of the sensing SINR versus the transmit power.
	As the transmit power gets higher, the sensing SINRs of the HBF-RadarOnly and HBF-Comm.Only designs are increasing, while the sensing SINRs of the FD-ISAC and the proposed HBF-ISAC designs remain around $20$dB across the considered range.
	This demonstrates that the proposed HBF-ISAC design can maintain a satisfactory sensing performance closer to that of the FD-ISAC design.

	\begin{figure}[!t]
		\centering  
		\includegraphics[width=0.75\linewidth]{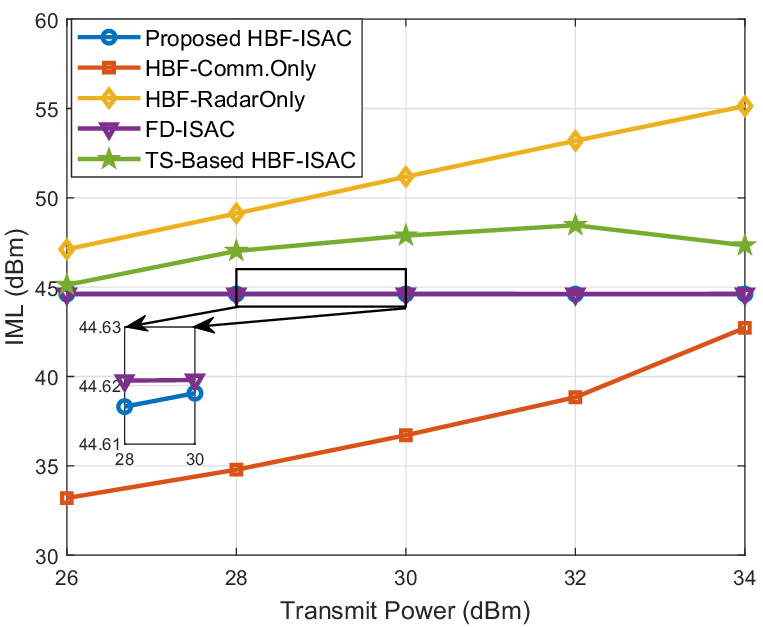}
		\vspace{-0.5em}
		\caption{Comparison of different schemes in terms of the IML versus transmit power.}
		\label{fig:7}
	\end{figure}
	
	To characterize the radiated power of the mainlobe region by the BS, we introduce a metric, integrated mainlobe level (IML), which is defined by
	\begin{equation}\label{eq:45}
		{\rm{IML = }}\sum\limits_{k = 1}^K {\sum\limits_{m = 1}^M {\left\| {{\bf{a}}_t^H\left( {{\theta _m},{f_k}} \right){{\bf{F}}_{{\rm{RF}}}}{{\bf{F}}_k}} \right\|_F^2} }.
	\end{equation}
	Fig. \ref{fig:7} plots curves of the IML versus the transmit power.
	We can observe that the IMLs realized by the proposed HBF-ISAC design are close, maintaining around $45$dBm with different transmit power, while the TS-based HBF-ISAC design causes an IML higher than $45$dBm.
	Besides, the HBF-RadarOnly design leads to the highest IML, even at $55$dBm with the transmit power $34$dBm.
	This verifies the effectiveness of the proposed HBF-ISAC design in terms of suppressing the radiated power to combat PD.

	\begin{figure*}[!t]
		\centering
		\subfigure[]{
			\includegraphics[width=0.195\linewidth]{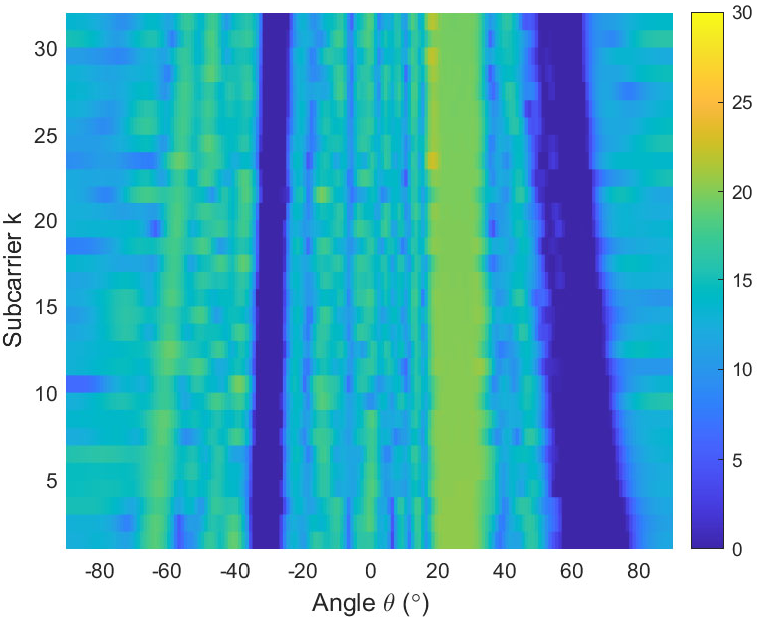} 
			\label{fig:8-1}	
			\hspace{-1em}}
		\subfigure[]{
			\includegraphics[width=0.195\linewidth]{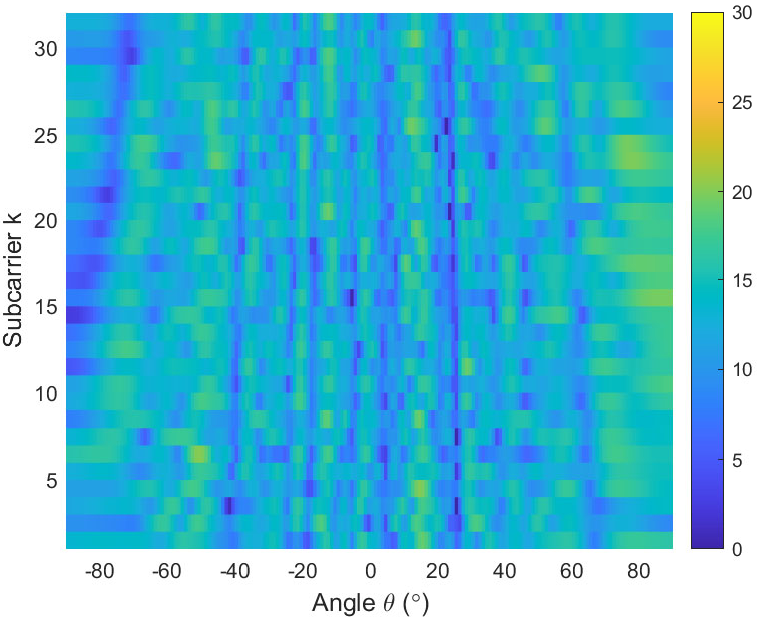} 
			\label{fig:8-2}
			\hspace{-1em}}
		\subfigure[]{
			\includegraphics[width=0.195\linewidth]{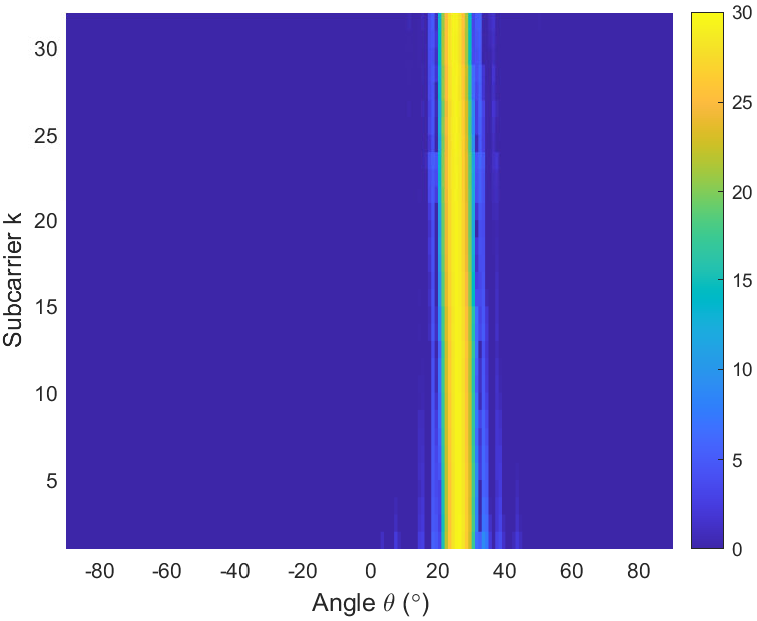} 
			\label{fig:8-3}
			\hspace{-1em}}
		\subfigure[]{
			\includegraphics[width=0.195\linewidth]{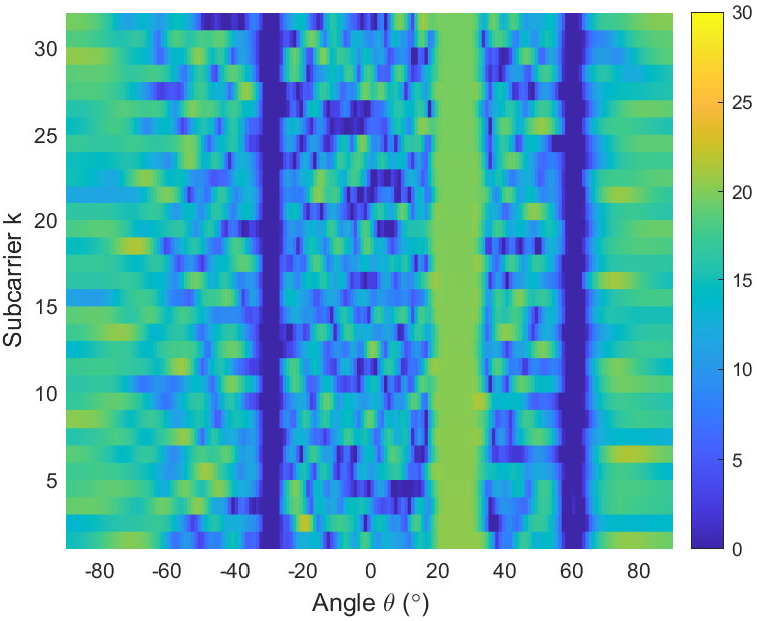} 
			\label{fig:8-4}
			\hspace{-1em}}
		\subfigure[]{
			\includegraphics[width=0.195\linewidth]{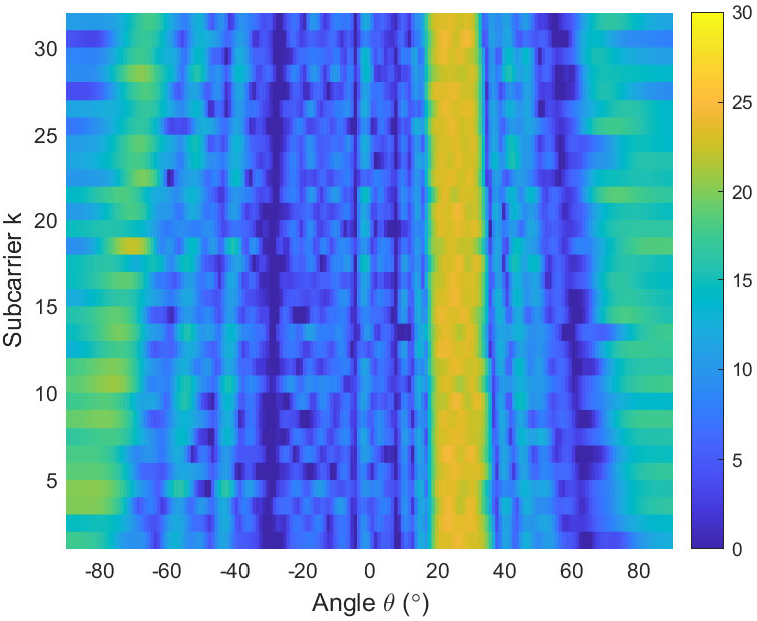} 
			\label{fig:8-5}}
		\vspace{-1em}
		\caption{Comparison of space-frequency spectra obtained by different design schemes. (a) Proposed HBF-ISAC design. (b) HBF-Comm.Only design. (c) HBF-RadarOnly design. (d) FD-ISAC design. (e) TS-based HBF-ISAC design.}
		\label{fig:8}
		\vspace{-0.5em}
	\end{figure*}
	
	Moreover, Fig. \ref{fig:8} compares the space-frequency spectra obtained by different design schemes.
	As can be seen from Fig. \ref{fig:8-1} and Fig. \ref{fig:8-4}, the space-frequency spectrum obtained by the proposed HBF-ISAC design and the FD-ISAC design can maintain $20$dBm in the mainlobe region and suppress the level in the clutter region to below $0$dBm.
	However, the space-frequency spectrum obtained by the TS-based HBF-ISAC design generates mainlobe level even higher than $25$dBm and fails to generate an extremely low nulling level toward the whole clutter region.
	This demonstrates that the proposed HBF-ISAC design can effectively achieve mainlobe control and clutter suppression, achieving space-frequency shaping performance comparable to the FD-ISAC scheme, while significantly outperforming the TS-based HBF-ISAC design.
	
	\begin{figure*}[!t]
		\centering
		\subfigure[]{
			\includegraphics[width=0.195\linewidth]{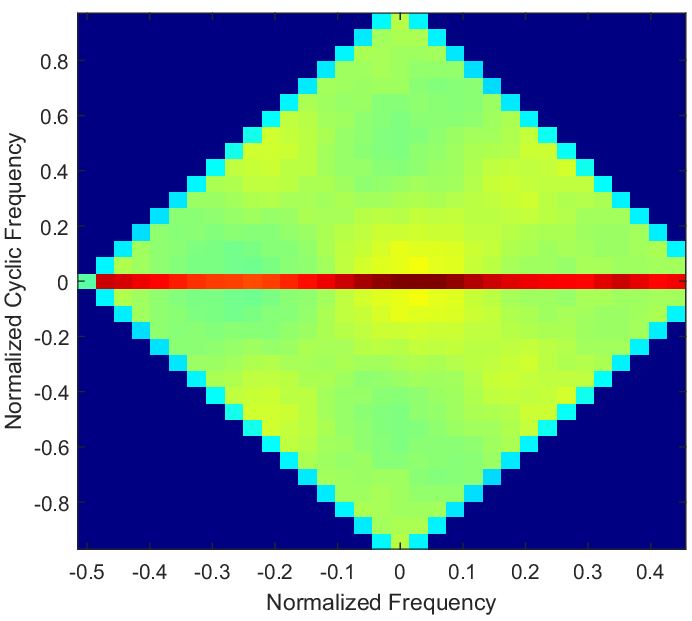} 
			\label{fig:9-1}	
			\hspace{-1em}}
		\subfigure[]{
			\includegraphics[width=0.195\linewidth]{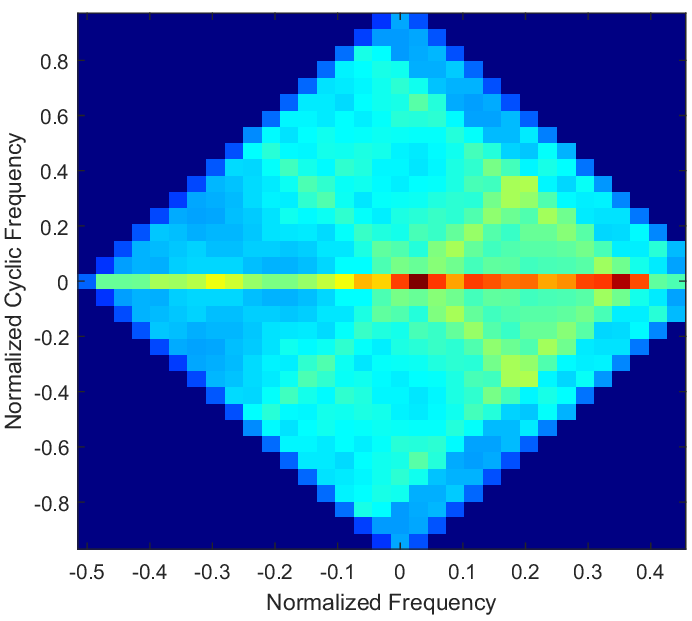} 
			\label{fig:9-2}
			\hspace{-1em}}
		\subfigure[]{
			\includegraphics[width=0.195\linewidth]{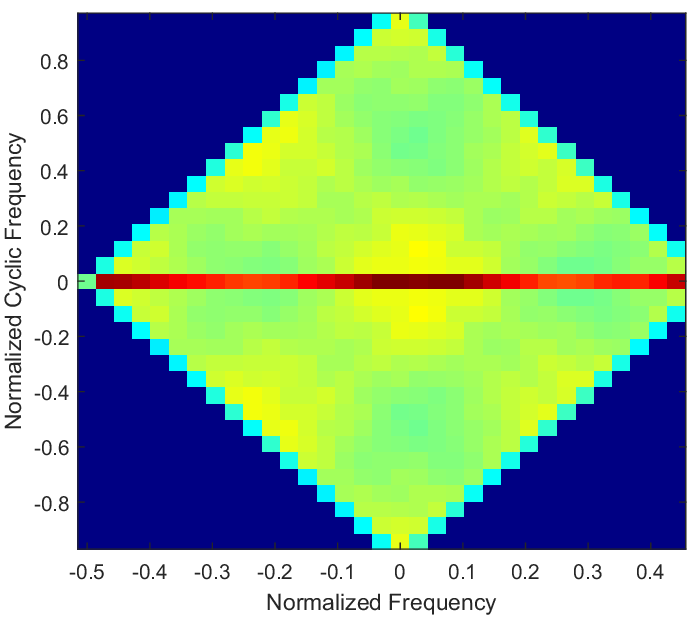} 
			\label{fig:9-3}
			\hspace{-1em}}
		\subfigure[]{
			\includegraphics[width=0.195\linewidth]{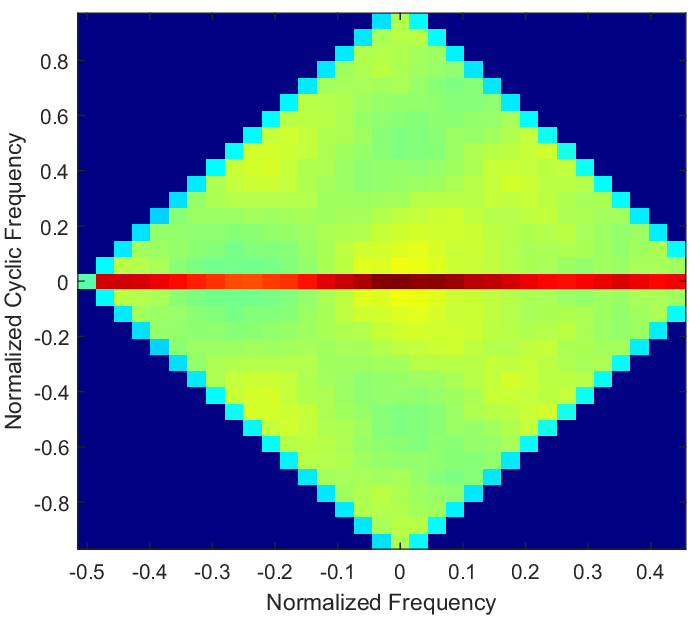} 
			\label{fig:9-4}
			\hspace{-1em}}
		\subfigure[]{
			\includegraphics[width=0.195\linewidth]{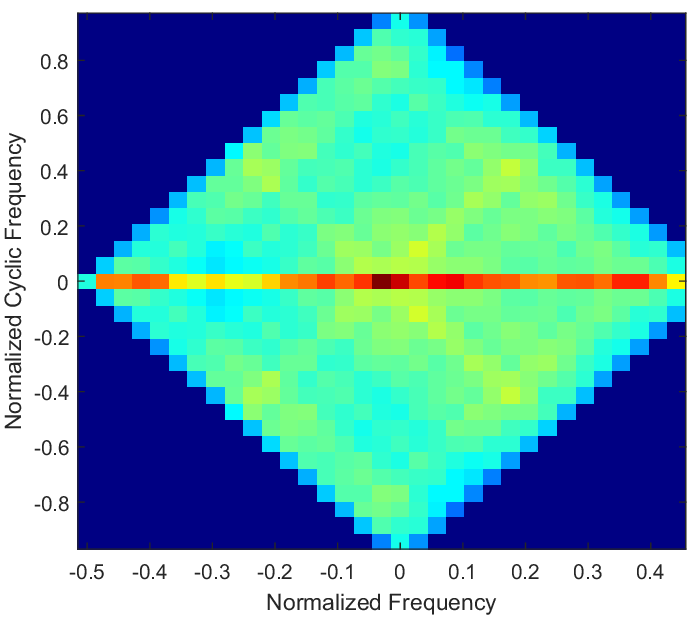} 
			\label{fig:9-5}}
		\vspace{-1em}
		\caption{Comparison of cyclic spectra obtained by different design schemes. (a) Proposed HBF-ISAC design. (b) HBF-Comm.Only design. (c) HBF-RadarOnly design. (d) FD-ISAC design. (e) TS-based HBF-ISAC design.}
		\label{fig:9}
	\end{figure*}
	
	Fig. \ref{fig:9} compares the cyclic spectra obtained by different design schemes, with the number of Monte Carlo trials is $N_{\rm MC} = 5000$.
	It can be seen from Fig. \ref{fig:9-1} and Fig. \ref{fig:9-4} that the cyclic spectra obtained by the proposed HBF-ISAC design and FD-ISAC design are similar to the cyclic spectrum of AWGN as shown in Fig. \ref{fig:CS-AWGN}.
	However, Fig. \ref{fig:9-2} and Fig. \ref{fig:9-5} show that the HBF-Comm.Only design and the TS-based HBF-ISAC design generate cyclic spectra obviously different from the cyclic spectrum of AWGN.
	This verifies the effectiveness of the proposed HBF-ISAC design in terms of generating noise-like cyclic spectrum to combat CA. 
	
	\subsubsection{Impact of the Number of RF Chains}
	
	\begin{figure}[!t]
		\centering  
		\includegraphics[width=0.75\linewidth]{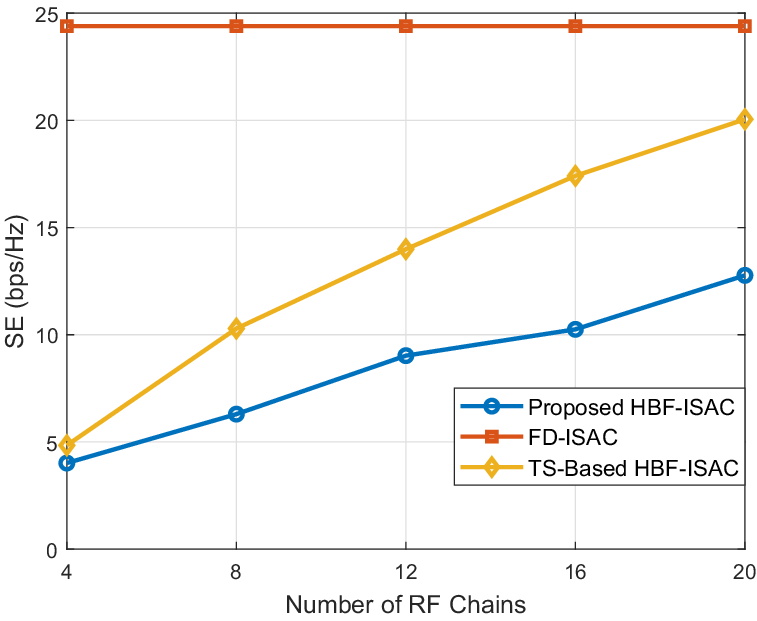}
		\vspace{-0.5em}
		\caption{The SE versus the number of RF chains.}
		\label{fig:10}
	\end{figure}
	
	\begin{figure}[!t]
		\centering  
		\includegraphics[width=0.75\linewidth]{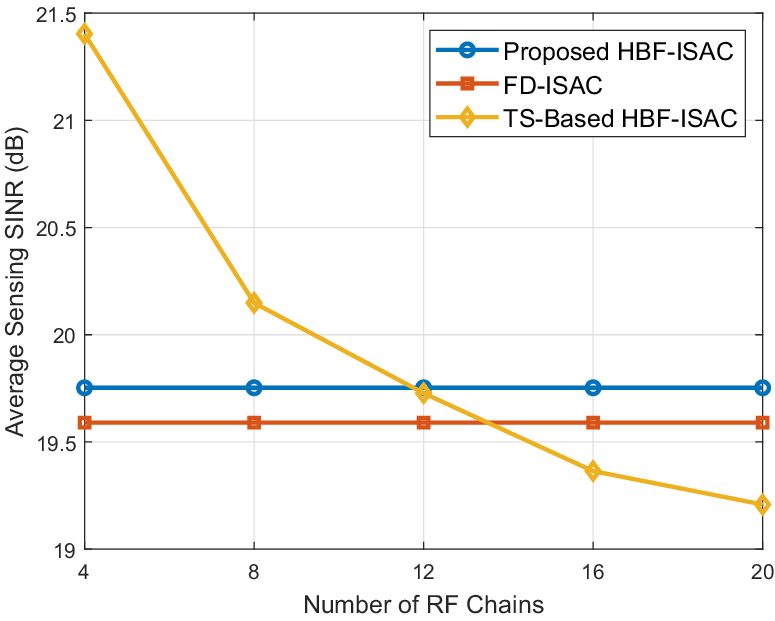}
		\vspace{-0.5em}
		\caption{The sensing SINR versus the number of RF chains.}
		\label{fig:11}
	\end{figure}
	
\begin{figure*}[!t]
		\centering
		\subfigure[]{
			\includegraphics[width=0.195\linewidth]{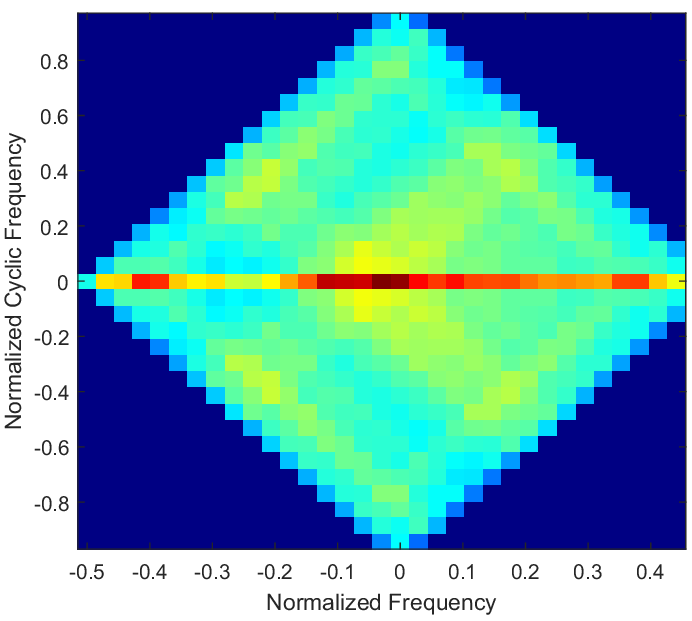} 
			\label{fig:12-1}	
			\hspace{-1em}}
		\subfigure[]{
			\includegraphics[width=0.195\linewidth]{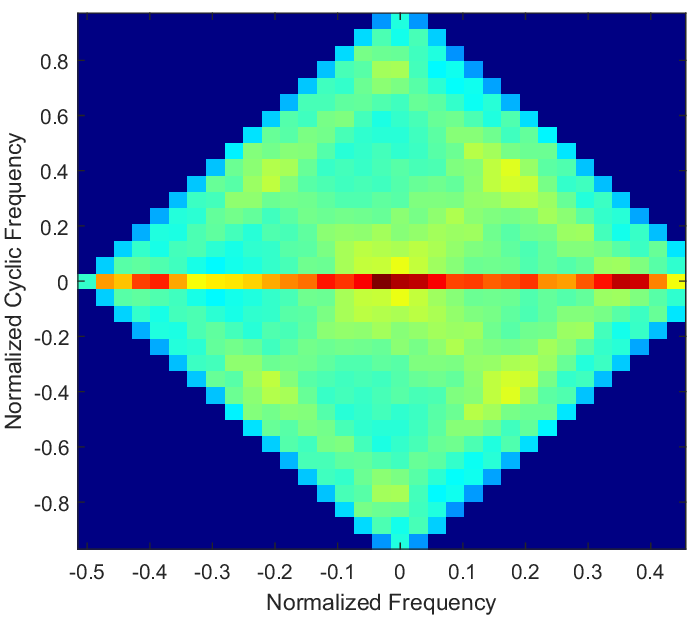} 
			\label{fig:12-2}
			\hspace{-1em}}
		\subfigure[]{
			\includegraphics[width=0.195\linewidth]{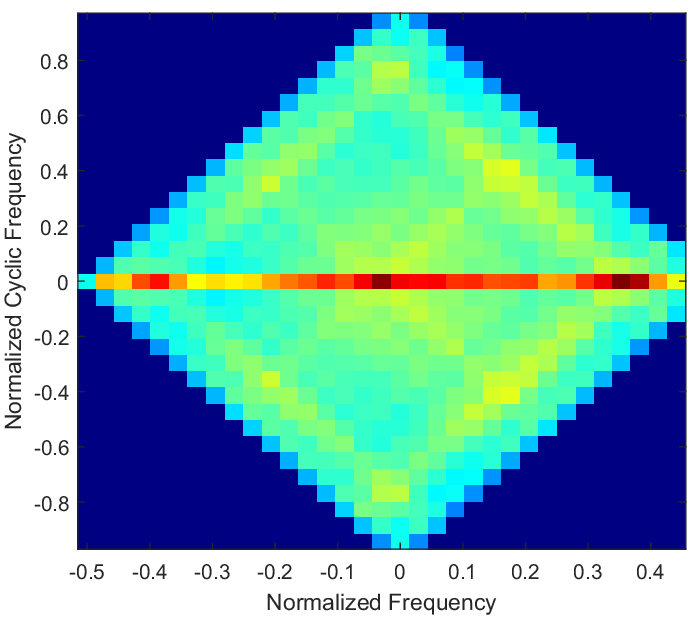} 
			\label{fig:12-3}
			\hspace{-1em}}
		\subfigure[]{
			\includegraphics[width=0.195\linewidth]{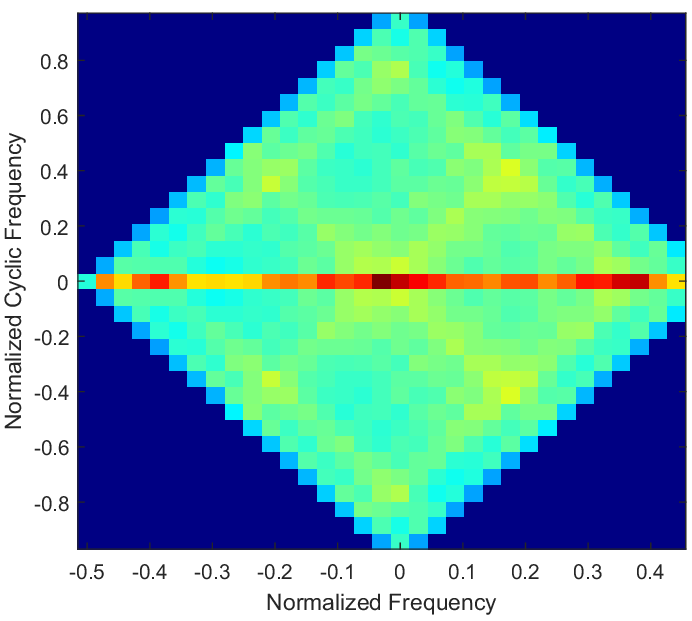} 
			\label{fig:12-4}
			\hspace{-1em}}
		\subfigure[]{
			\includegraphics[width=0.195\linewidth]{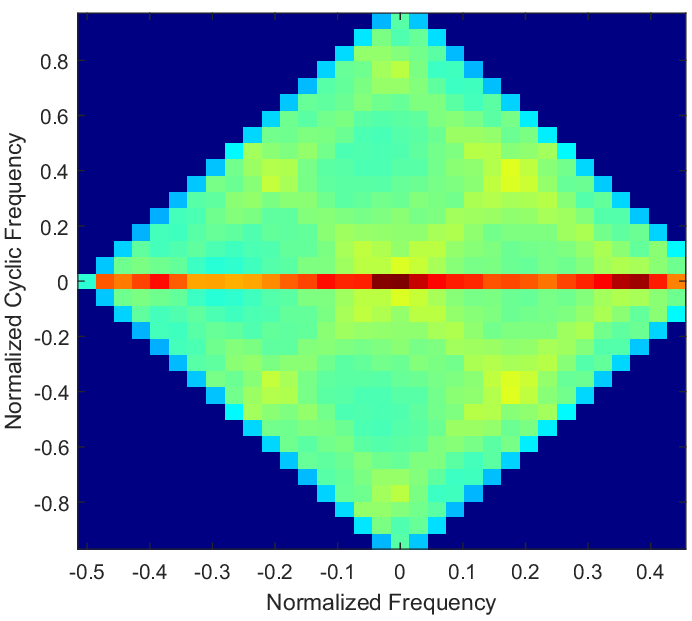} 
			\label{fig:12-5}}
		\vspace{-1em}
		\caption{Comparison of cyclic spectra obtained by TS-based HBF-ISAC schemes with a different number of RF chains. (a) $N_{\rm RF} = 4$. (b) $N_{\rm RF} = 8$. (c) $N_{\rm RF} = 12$. (d) $N_{\rm RF} = 16$. (e) $N_{\rm RF} = 20$.}
		\label{fig:12}
		\vspace{-0.5em}
	\end{figure*}

\begin{figure*}[!t]
		\centering
		\subfigure[]{
			\includegraphics[width=0.195\linewidth]{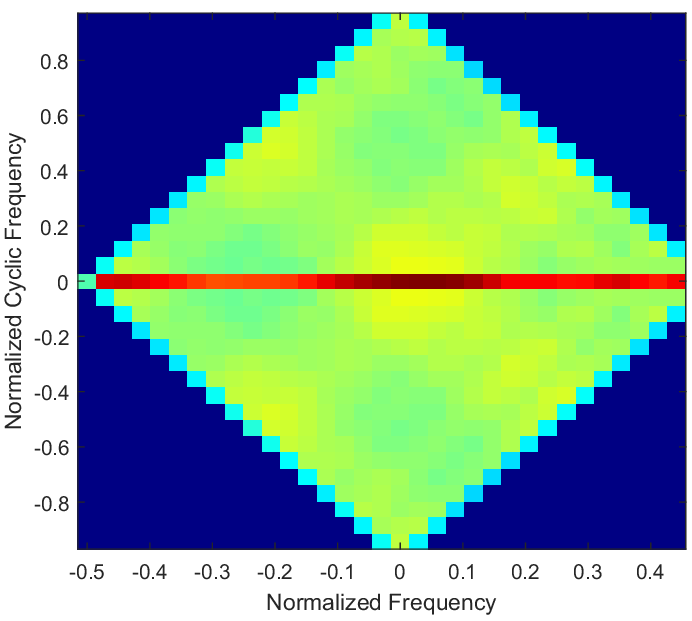} 
			\label{fig:13-1}	
			\hspace{-1em}}
		\subfigure[]{
			\includegraphics[width=0.195\linewidth]{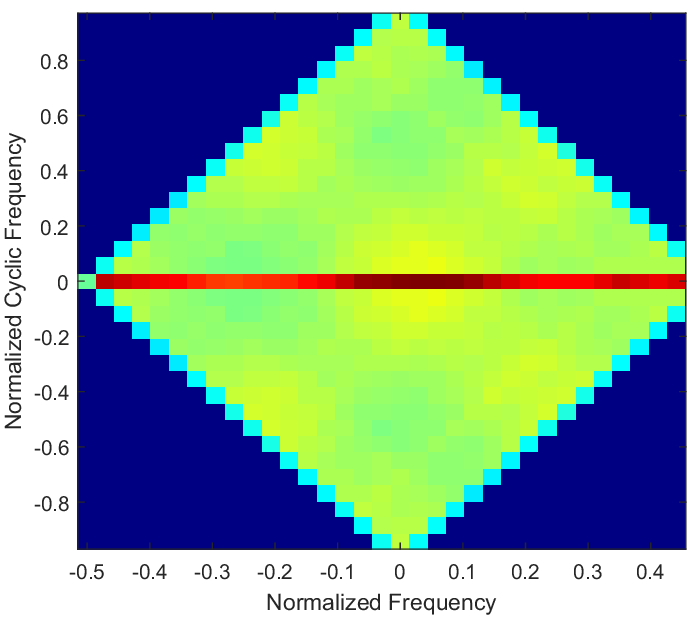} 
			\label{fig:13-2}
			\hspace{-1em}}
		\subfigure[]{
			\includegraphics[width=0.195\linewidth]{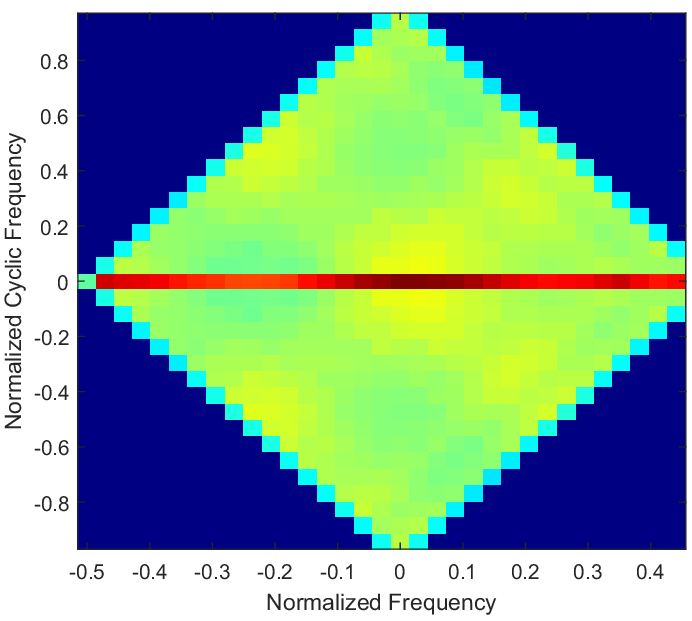} 
			\label{fig:13-3}
			\hspace{-1em}}
		\subfigure[]{
			\includegraphics[width=0.195\linewidth]{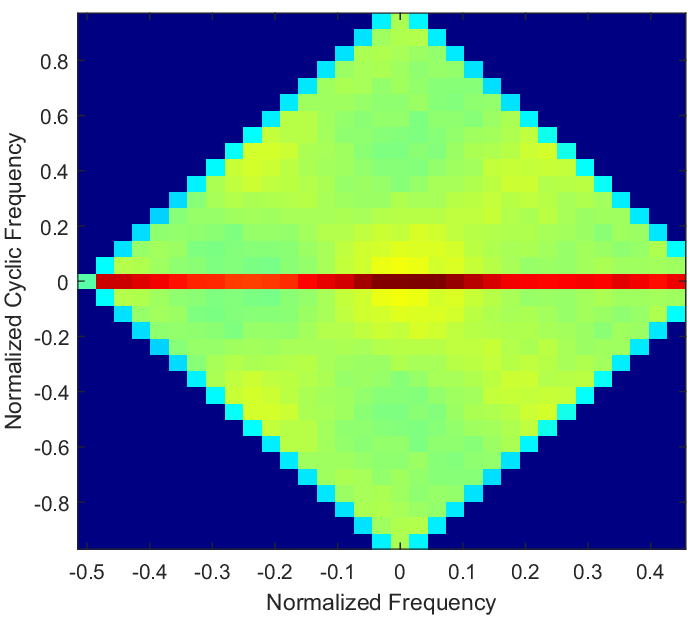} 
			\label{fig:13-4}
			\hspace{-1em}}
		\subfigure[]{
			\includegraphics[width=0.195\linewidth]{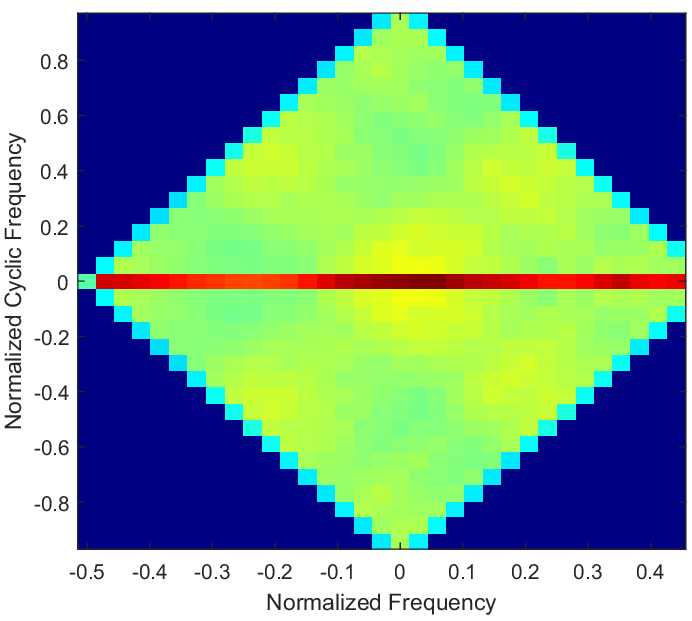} 
			\label{fig:13-5}}
		\vspace{-1em}
		\caption{Comparison of cyclic spectra obtained by the proposed HBF-ISAC schemes with a different number of RF chains. (a) $N_{\rm RF} = 4$. (b) $N_{\rm RF} = 8$. (c) $N_{\rm RF} = 12$. (d) $N_{\rm RF} = 16$. (e) $N_{\rm RF} = 20$.}
		\label{fig:13}
	\end{figure*}

	Furthermore, we evaluate the impact of the number of RF chains on the performance.
	Fig. \ref{fig:10} plots the curves of the SE versus the number of RF chains.
	It can be seen that the SE achieved by the proposed HBF-ISAC and the TS-based HBF-ISAC designs increases as the number of RF chains increases.
	Besides, Fig. \ref{fig:11} plots the curves of the sensing SINR versus the number of RF chains.
	As the number of RF chains increases, the sensing SINR realized by the proposed HBF-ISAC design increases and approaches that realized by the FD-ISAC design, while the sensing SINR realized by the TS-based HBF-ISAC design decreases.
	Moreover, Figs. \ref{fig:12} and \ref{fig:13} compare the cyclic spectra obtained by the TS-based HBF-ISAC design and the proposed HBF-ISAC design, respectively.
	It can be seen that the cyclic spectra obtained by the proposed HBF-ISAC design are similar to that of the AWGN with a different number of RF chains, while those obtained by the TS-based HBF-ISAC design are different from that of the AWGN.
	This indicates that a suitable choice of the RF chain number settings can improve the communication and sensing performance.
	It also validates the superiority of the proposed HBF-ISAC design over the TS-based HBF-ISAC design in terms of sensing security performance.

	\subsubsection{Impact of the Number of Subcarriers}
	
\begin{figure}[!t]
		\centering  
		\includegraphics[width=0.75\linewidth]{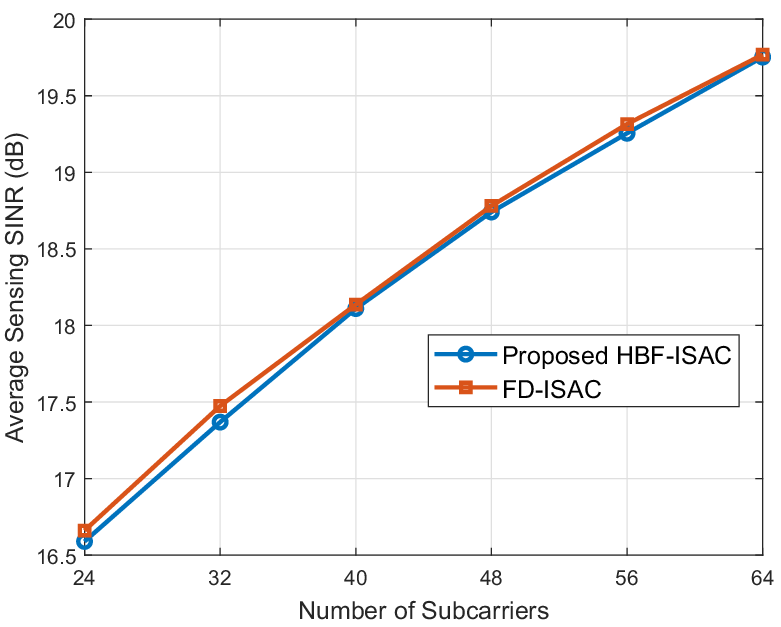}
		\vspace{-0.5em}
		\caption{The sensing SINR versus the number of subcarriers.}
		\label{fig:14}
	\end{figure}

	Furthermore, Fig. \ref{fig:14} plots the sensing SINR versus the number of subcarriers of the proposed HBF-ISAC and FD-ISAC design.
	It can be seen that the sensing SINR is higher with a higher number of subcarriers, and the sensing SINR obtained by the proposed HBF-ISAC design is close to that of the FD-ISAC design.
	This validates the effectiveness of the proposed design and algorithm with different numbers of subcarriers in the OFDM-ISAC systems.
	
	\subsubsection{Impact of Imperfect CSI}
	
\begin{figure}[!t]
		\centering  
		\includegraphics[width=0.75\linewidth]{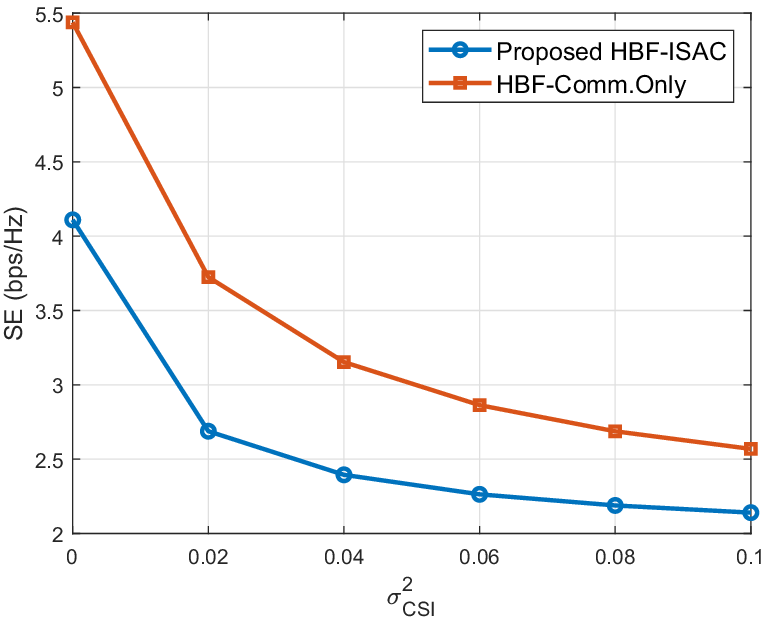}
		\vspace{-0.5em}
		\caption{The SE versus the estimated CSI error.}
		\label{fig:15}
	\end{figure}

	To further evaluate the impact of imperfect CSI, we performed simulations under an imperfect CSI model, where the communication channel is ${\bf h} = \hat{\bf h} + \Delta{\bf h}$, with $\hat{\bf h}$ denoting the estimate CSI and $\Delta{\bf h}\sim{\cal CN}(0,\sigma_{\rm CSI}^2{\bf I}_{M_t})$ denotes the estimate error.
	Fig. \ref{fig:15} plots the SE versus the estimate CSI error of the proposed HBF-ISAC and HBF-Comm.Only design.
	It can be seen that the SE decreases as the estimate CSI error increases, with moderate performance degradation.
	Besides, the SE degradations of the proposed design and the HBF-Comm.Only are similar.
	This indicates that the proposed design is reasonably robust to practical CSI imperfections.
	
	\section{Conclusion}
	
	In this paper, we investigated improving the sensing security of OFDM-ISAC systems in the presence of an advanced ER aircraft that employs both PD and CA intercept methods.
	We first analyzed the power distribution and cyclic spectrum of ISAC signals, which are vital parameters involved in the decision-making mechanism for PD and CA, respectively.
	Specifically, we analyzed the intrinsic mathematical properties of a newly introduced metric, ergodic cyclic spectrum, to characterize the cyclostationary behavior.
	To enhance sensing security for ISAC signals, we devised an LPI-aware HBF design by maximizing the communication SE subject to the probability of intercept, sensing, power, and hardware constraints.
	To deal with the uncertainty of the target location and the unavailability of a specific ER decision-making mechanism, we devised a robust LPI design scheme and proposed an effective algorithm based on an alternating optimization approach.
	Numerical simulations verify the convergence and effectiveness of the proposed algorithm.
	Moreover, it demonstrates that the proposed LPI-aware HBF design for the OFDM-ISAC system can enhance the system security against multi-intercept threats while guaranteeing satisfactory radar detection and communication performance.
	In future work, the joint consideration of sensing security and communication security in ISAC systems and robust design without relying on instantaneous CSI will be investigated as a valuable extension of this work.
	
	\appendices

	\section{Proof of Lemma \ref{lemma:1}}\label{app:1}
	
	Substituting ${f_k} = {f_c} + \left( {k - \frac{K}{2}} \right)\Delta f$ and $T_s = \frac{1}{N \Delta f}$ into the ergodic cyclic spectrum \eqref{eq:17}, we obtain
	\begin{equation}\label{eq:A1}
		{{\bf{R}}_\Xi }\left[ {m,n} \right] = e^{\jmath {2}\pi \left( {{f_c} - \frac{K}{2}\Delta f} \right)\left( {m - n} \right){T_s}}\sum\limits_{k = 1}^K c_k^2{{e^{\jmath {2}\pi k\frac{{m - n}}{N}}}} .
	\end{equation}
	Considering the complex exponential summation
	\begin{equation}\label{eq:A2}
		\sum\limits_{k = 1}^K {{e^{\jmath {2}\pi k\frac{{m - n}}{N}}}} = {e^{\jmath \pi \left( {K + 1} \right)\frac{{m - n}}{N}}}\frac{{\sin \left( {\pi K\frac{{m - n}}{N}} \right)}}{{\sin \left( {\pi \frac{{m - n}}{N}} \right)}},
	\end{equation}
	if $c_k^2 = c^2, \forall k$, the magnitude of the ergodic cyclic spectrum \eqref{eq:A1} can be expressed as
	\begin{equation}\label{eq:A3}
		\left| {{{\bf{R}}_\Xi }\left[ {m,n} \right]} \right| = {c^2}\left| {\frac{{\sin \left( {\pi K\frac{{m - n}}{N}} \right)}}{{\sin \left( {\pi \frac{{m - n}}{N}} \right)}}} \right|.
	\end{equation}
	Based on \eqref{eq:A3}, we consider the following two cases
	\begin{itemize}
		\item Case 1: For the diagonal entries, namely $m = n$, using the limit ${\lim _{x \to 0}}\frac{{\sin \left( {\pi Kx} \right)}}{{\sin \left( {\pi x} \right)}} = K$, we have 
		\begin{equation}
			\left| {{{\bf{R}}_\Xi }\left[ {m,n} \right]} \right| = K{c^2}.
		\end{equation}
		\item Case 2: For the off-diagonal entries, namely $m \ne n$, the magnitude of the Dirichlet kernel $\frac{{\sin \left( {\pi K\frac{{m - n}}{N}} \right)}}{{\sin \left( {\pi \frac{{m - n}}{N}} \right)}}$ is significantly smaller than the main peak, and in most positions approaches zero, with exact zeros occurring at periodic intervals.
	\end{itemize}
	Therefore, after normalizing by the main diagonal, the matrix exhibits a predominantly diagonal structure:
	\begin{equation}
		\frac{1}{{{c^2}K}} {{{\bf{R}}_\Xi }} \approx {{\bf{I}}_N}.
	\end{equation}
	
	The proof is completed.

	\balance
	\bibliographystyle{IEEEtran}
	\bibliography{IEEEabrv,ref_abrv.bib}

\begin{thebibliography}{10}
\providecommand{\url}[1]{#1}
\csname url@samestyle\endcsname
\providecommand{\newblock}{\relax}
\providecommand{\bibinfo}[2]{#2}
\providecommand{\BIBentrySTDinterwordspacing}{\spaceskip=0pt\relax}
\providecommand{\BIBentryALTinterwordstretchfactor}{4}
\providecommand{\BIBentryALTinterwordspacing}{\spaceskip=\fontdimen2\font plus
\BIBentryALTinterwordstretchfactor\fontdimen3\font minus
  \fontdimen4\font\relax}
\providecommand{\BIBforeignlanguage}[2]{{%
\expandafter\ifx\csname l@#1\endcsname\relax
\typeout{** WARNING: IEEEtran.bst: No hyphenation pattern has been}%
\typeout{** loaded for the language `#1'. Using the pattern for}%
\typeout{** the default language instead.}%
\else
\language=\csname l@#1\endcsname
\fi
#2}}
\providecommand{\BIBdecl}{\relax}
\BIBdecl

\bibitem{liu2022integrated}
F.~Liu, Y.~Cui, C.~Masouros \emph{et~al.}, ``{Integrated sensing and
  communications: Toward dual-functional wireless networks for 6G and
  beyond},'' \emph{{IEEE} J. Sel. Areas Commun.}, vol.~40, no.~6, pp.
  1728--1767, 2022.

\bibitem{wei2023integrated}
Z.~Wei, H.~Qu, Y.~Wang \emph{et~al.}, ``{Integrated sensing and communication
  signals toward 5G-A and 6G: A survey},'' \emph{{IEEE} Internet Things J.},
  vol.~10, no.~13, pp. 11\,068--11\,092, 2023.

\bibitem{cui2021integrating}
Y.~Cui, F.~Liu, X.~Jing, and J.~Mu, ``{Integrating sensing and communications
  for ubiquitous IoT: Applications, trends, and challenges},'' \emph{{IEEE}
  Netw.}, vol.~35, no.~5, pp. 158--167, 2021.

\bibitem{saad2019vision}
W.~Saad, M.~Bennis, and M.~Chen, ``{A vision of 6G wireless systems:
  Applications, trends, technologies, and open research problems},''
  \emph{{IEEE} Netw.}, vol.~34, no.~3, pp. 134--142, 2019.

\bibitem{wang2023road}
C.-X. Wang, X.~You, X.~Gao \emph{et~al.}, ``{On the road to 6G: Visions,
  requirements, key technologies, and testbeds},'' \emph{{IEEE} Commun. Surveys
  Tuts.}, vol.~25, no.~2, pp. 905--974, 2023.

\bibitem{giordani2020toward}
M.~Giordani, M.~Polese, M.~Mezzavilla, S.~Rangan, and M.~Zorzi, ``{Toward 6G
  networks: Use cases and technologies},'' \emph{{IEEE} Commun. Mag.}, vol.~58,
  no.~3, pp. 55--61, 2020.

\bibitem{cheng2022lightweight}
Y.~Cheng, J.~Ma, Z.~Liu, Y.~Wu, K.~Wei, and C.~Dong, ``{A lightweight privacy
  preservation scheme with efficient reputation management for mobile
  crowdsensing in vehicular networks},'' \emph{{IEEE} Trans. Depend. Sec.
  Comput.}, vol.~20, no.~3, pp. 1771--1788, 2022.

\bibitem{zou2016survey}
Y.~Zou, J.~Zhu, X.~Wang, and L.~Hanzo, ``{A survey on wireless security:
  Technical challenges, recent advances, and future trends},'' \emph{Proc.
  {IEEE}}, vol. 104, no.~9, pp. 1727--1765, 2016.

\bibitem{shiu2011physical}
Y.-S. Shiu, S.~Y. Chang, H.-C. Wu, S.~C.-H. Huang, and H.-H. Chen, ``{Physical
  layer security in wireless networks: A tutorial},'' \emph{{IEEE} Wireless
  Commun.}, vol.~18, no.~2, pp. 66--74, 2011.

\bibitem{mukherjee2014principles}
A.~Mukherjee, S.~A.~A. Fakoorian, J.~Huang, and A.~L. Swindlehurst,
  ``{Principles of physical layer security in multiuser wireless networks: A
  survey},'' \emph{{IEEE} Commun. Surveys Tuts.}, vol.~16, no.~3, pp.
  1550--1573, 2014.

\bibitem{liu2016physical}
Y.~Liu, H.-H. Chen, and L.~Wang, ``{Physical layer security for next generation
  wireless networks: Theories, technologies, and challenges},'' \emph{{IEEE}
  Commun. Surveys Tuts.}, vol.~19, no.~1, pp. 347--376, 2016.

\bibitem{wei2022toward}
Z.~Wei, F.~Liu, C.~Masouros, N.~Su, and A.~P. Petropulu, ``{Toward
  multi-functional 6G wireless networks: Integrating sensing, communication,
  and security},'' \emph{{IEEE} Commun. Mag.}, vol.~60, no.~4, pp. 65--71,
  2022.

\bibitem{jiang2024physical}
Y.~Jiang, L.~Wang, H.-H. Chen, and X.~Shen, ``{Physical Layer Covert
  Communication in B5G Wireless Networks—its Research, Applications, and
  Challenges},'' \emph{Proc. {IEEE}}, vol. 112, no.~1, pp. 47--82, 2024.

\bibitem{chu2023joint}
J.~Chu, Z.~Lu, R.~Liu, M.~Li, and Q.~Liu, ``{Joint beamforming and reflection
  design for secure RIS-ISAC systems},'' \emph{{IEEE} Trans. Veh. Technol.},
  vol.~73, no.~3, pp. 4471--4475, 2023.

\bibitem{xu2022robust}
D.~Xu, X.~Yu, D.~W.~K. Ng, A.~Schmeink, and R.~Schober, ``{Robust and secure
  resource allocation for ISAC systems: A novel optimization framework for
  variable-length snapshots},'' \emph{{IEEE} Trans. Commun.}, vol.~70, no.~12,
  pp. 8196--8214, 2022.

\bibitem{ren2023robust}
Z.~Ren, L.~Qiu, J.~Xu, and D.~W.~K. Ng, ``{Robust transmit beamforming for
  secure integrated sensing and communication},'' \emph{{IEEE} Trans. Commun.},
  vol.~71, no.~9, pp. 5549--5564, 2023.

\bibitem{wang2024sensing}
X.~Wang, Z.~Fei, P.~Liu, J.~A. Zhang, Q.~Wu, and N.~Wu, ``{Sensing-aided covert
  communications: Turning interference into allies},'' \emph{{IEEE} Trans.
  Wireless Commun.}, vol.~23, no.~9, pp. 10\,726--10\,739, 2024.

\bibitem{su2020secure}
N.~Su, F.~Liu, and C.~Masouros, ``{Secure radar-communication systems with
  malicious targets: Integrating radar, communications and jamming
  functionalities},'' \emph{{IEEE} Trans. Wireless Commun.}, vol.~20, no.~1,
  pp. 83--95, 2020.

\bibitem{guo2024secure}
L.~Guo, J.~Jia, J.~Chen, S.~Yang, and X.~Wang, ``{Secure Beamforming and Radar
  Association in CoMP-NOMA Empowered Integrated Sensing and Communication
  Systems},'' \emph{{IEEE} Trans. Inf. Forensics Security}, vol.~19, pp.
  10\,246--10\,257, 2024.

\bibitem{liu2022outage}
P.~Liu, Z.~Fei, X.~Wang, B.~Li, Y.~Huang, and Z.~Zhang, ``{Outage constrained
  robust secure beamforming in integrated sensing and communication systems},''
  \emph{{IEEE} Wireless Commun. Lett.}, vol.~11, no.~11, pp. 2260--2264, 2022.

\bibitem{hu2017cooperative}
L.~Hu, H.~Wen, B.~Wu \emph{et~al.}, ``{Cooperative Jamming for Physical Layer
  Security Enhancement in Internet of Things},'' \emph{{IEEE} Internet Things
  J.}, vol.~5, no.~1, pp. 219--228, 2018.

\bibitem{gu2023robust}
J.~Gu, G.~Ding, H.~Wang, and Y.~Xu, ``{Robust beamforming for sensing assisted
  integrated communication and jamming systems},'' \emph{{IEEE} Trans. Veh.
  Technol.}, vol.~72, no.~10, pp. 13\,649--13\,654, 2023.

\bibitem{ye2023robust}
R.~Ye, Y.~Peng, F.~Al-Hazemi, and R.~Boutaba, ``{A robust cooperative jamming
  scheme for secure UAV communication via intelligent reflecting surface},''
  \emph{{IEEE} Trans. Commun.}, vol.~72, no.~2, pp. 1005--1019, 2023.

\bibitem{magbool2025survey}
A.~Magbool, V.~Kumar, Q.~Wu, M.~Di~Renzo, and M.~F. Flanagan, ``{A Survey on
  Integrated Sensing and Communication With Intelligent Metasurfaces: Trends,
  Challenges, and Opportunities},'' \emph{IEEE Open J. Commun. Soc.}, vol.~6,
  pp. 7270--7318, 2025.

\bibitem{kumar2025beamforming}
V.~Kumar and M.~Chafii, ``{Beamforming Design for Secure RIS-Enabled ISAC:
  Passive RIS Versus Active RIS},'' \emph{{IEEE} Trans. Wireless Commun.},
  vol.~24, no.~9, pp. 7719--7732, 2025.

\bibitem{kazymova2025achievable}
A.~Kazymova, V.~Kumar, C.~Pöpper, and M.~Chafii, ``{Achievable Sum Secrecy
  Rate of STAR-RIS-Enabled MU-MIMO ISAC},'' in \emph{IEEE ICC Workshops 2025},
  2025, pp. 947--952.

\bibitem{10643599}
B.~Wang, H.~Li, S.~Shen, Z.~Cheng, and B.~Clerckx, ``{A Dual-Function
  Radar-Communication System Empowered by Beyond Diagonal Reconfigurable
  Intelligent Surface},'' \emph{{IEEE} Trans. Commun.}, vol.~73, no.~3, pp.
  1501--1516, 2025.

\bibitem{montezuma2015implementing}
P.~Montezuma and R.~Dinis, ``{Implementing physical layer security using
  transmitters with constellation shaping},'' in \emph{ICCCN 2015}.\hskip 1em
  plus 0.5em minus 0.4em\relax IEEE, 2015, pp. 1--4.

\bibitem{bang2020secure}
I.~Bang and T.~Kim, ``{Secure modulation based on constellation mapping
  obfuscation in OFDM based TDD systems},'' \emph{{IEEE} Access}, vol.~8, pp.
  197\,644--197\,653, 2020.

\bibitem{ma2022optimal}
S.~Ma, Y.~Zhang, H.~Sheng \emph{et~al.}, ``{Optimal probabilistic constellation
  shaping for covert communications},'' \emph{{IEEE} Trans. Inf. Forensics
  Security}, vol.~17, pp. 3165--3178, 2022.

\bibitem{stove2004low}
A.~Stove, A.~Hume, and C.~Baker, ``{Low probability of intercept radar
  strategies},'' \emph{IEE Proceedings-Radar, Sonar and Navigation}, vol. 151,
  no.~5, pp. 249--260, 2004.

\bibitem{pace2009detecting}
P.~E. Pace, \emph{{Detecting and classifying low probability of intercept
  radar}}.\hskip 1em plus 0.5em minus 0.4em\relax Artech house, 2009.

\bibitem{liu2015lpi}
Y.~Liu, P.~Xiao, H.~Wu \emph{et~al.}, ``{LPI radar signal detection based on
  radial integration of Choi-Williams time-frequency image},'' \emph{Journal of
  systems engineering and electronics}, vol.~26, no.~5, pp. 973--981, 2015.

\bibitem{zhang2016lpi}
M.~Zhang, L.~Liu, and M.~Diao, ``{LPI radar waveform recognition based on
  time-frequency distribution},'' \emph{Sensors}, vol.~16, no.~10, p. 1682,
  2016.

\bibitem{magbool2025hiding}
A.~Magbool, V.~Kumar, M.~D. Renzo, and M.~F. Flanagan, ``{Hiding in Plain
  Sight: RIS-Aided Target Obfuscation in ISAC},'' \emph{{IEEE} Trans. Wireless
  Commun.}, vol.~25, pp. 14\,550--14\,563, 2026.

\bibitem{han2025sensing}
K.~Han, K.~Meng, and C.~Masouros, ``{Sensing-Secure ISAC: Ambiguity Function
  Engineering for Impairing Unauthorized Sensing},'' \emph{{IEEE} Trans.
  Wireless Commun.}, vol.~25, pp. 5386--5400, 2026.

\bibitem{rexhepi2025blinding}
G.~Rexhepi, H.~S. Rou, G.~Thadeu Freitas~de Abreu, and G.~C. Alexandropoulos,
  ``{Blinding the Wiretapper: RIS-Enabled User Occultation in the ISAC Era},''
  in \emph{Asilomar 2025}, 2025, pp. 749--753.

\bibitem{shi2018low}
C.~Shi, F.~Wang, S.~Salous, and J.~Zhou, ``{Low probability of intercept-based
  optimal OFDM waveform design strategy for an integrated radar and
  communications system},'' \emph{{IEEE} Access}, vol.~6, pp. 57\,689--57\,699,
  2018.

\bibitem{shi2019low}
C.~Shi, F.~Wang, M.~Sellathurai, J.~Zhou, and S.~Salous, ``{Low probability of
  intercept-based optimal power allocation scheme for an integrated multistatic
  radar and communication system},'' \emph{{IEEE} Syst. J.}, vol.~14, no.~1,
  pp. 983--994, 2019.

\bibitem{wang2020lpi}
Y.~Wang, C.~Shi, F.~Wang, and J.~Zhou, ``{LPI-based optimal radar power
  allocation for target time delay estimation in joint radar and communications
  system},'' in \emph{SAM 2020}, 2020, pp. 1--4.

\bibitem{gong2022joint}
P.~Gong, Z.~Zhang, Y.~Wu, and W.-Q. Wang, ``{Joint design of transmit waveform
  and receive beamforming for LPI FDA-MIMO radar},'' \emph{{IEEE} Signal
  Process. Lett.}, vol.~29, pp. 1938--1942, 2022.

\bibitem{gupta2019feature}
A.~Gupta and A.~A. Bazil~Rai, ``{Feature Extraction of Intra-Pulse Modulated
  LPI Waveforms Using STFT},'' in \emph{RTEICT 2019}, 2019, pp. 742--746.

\bibitem{zilberman2006autonomous}
E.~R. Zilberman and P.~E. Pace, ``{Autonomous time-frequency morphological
  feature extraction algorithm for LPI radar modulation classification},'' in
  \emph{ICIP 2006}, 2006, pp. 2321--2324.

\bibitem{chilukuri2020estimation}
R.~K. Chilukuri, H.~K. Kakarla, and K.~Subbarao, ``{Estimation of modulation
  parameters of LPI radar using cyclostationary method},'' \emph{Sensing and
  Imaging}, vol.~21, no.~1, p.~51, 2020.

\bibitem{liu2023integrated}
X.~Liu, Y.~Yuan, T.~Zhang \emph{et~al.}, ``{Integrated transmit waveform and
  RIS phase shift design for LPI detection and communication},'' \emph{{IEEE}
  Trans. Wireless Commun.}, vol.~23, no.~6, pp. 5663--5679, 2023.

\bibitem{shi2025low}
Q.~Shi, Y.~Wang, Z.~Zhou, G.~Cui, and P.~Fan, ``{Low Probability of Intercept
  Signal Design for MIMO Integrated Sensing and Communication Systems},''
  \emph{{IEEE} Trans. Commun.}, vol.~73, no.~9, pp. 8155--8165, 2025.

\bibitem{alkhateeb2014channel}
A.~Alkhateeb, O.~El~Ayach, G.~Leus, and R.~W. Heath, ``{Channel estimation and
  hybrid precoding for millimeter wave cellular systems},'' \emph{{IEEE} J.
  Sel. Topics Signal Process.}, vol.~8, no.~5, pp. 831--846, 2014.

\bibitem{11363235}
B.~Wang, H.~Li, F.~Liu, Z.~Cheng, and S.~Shen, ``{Distributed Hybrid
  Beamforming Design for Cooperative Cell-Free Integrated Sensing and
  Communication Networks},'' \emph{{IEEE} Trans. Commun.}, vol.~74, pp.
  4203--4219, 2026.

\bibitem{wang2022partially}
X.~Wang, Z.~Fei, J.~A. Zhang, and J.~Xu, ``{Partially-connected hybrid
  beamforming design for integrated sensing and communication systems},''
  \emph{{IEEE} Trans. Commun.}, vol.~70, no.~10, pp. 6648--6660, 2022.

\bibitem{sturm2009novel}
C.~Sturm, E.~Pancera, T.~Zwick, and W.~Wiesbeck, ``{A novel approach to OFDM
  radar processing},'' in \emph{IEEE RadarConf 2009}, 2009, pp. 1--4.

\bibitem{liu2019evaluation}
Y.~Liu, J.~Yi, X.~Wan, X.~Zhang, and H.~Ke, ``{Evaluation of clutter
  suppression in CP-OFDM-based passive radar},'' \emph{{IEEE} Sensors J.},
  vol.~19, no.~14, pp. 5572--5586, 2019.

\bibitem{liu2020time}
------, ``{Time-varying clutter suppression in CP-OFDM based passive radar for
  slowly moving targets detection},'' \emph{{IEEE} Sensors J.}, vol.~20,
  no.~16, pp. 9079--9090, 2020.

\bibitem{10858124}
W.~Deng, M.~Li, M.-M. Zhao \emph{et~al.}, ``{CSI Transfer From Sub-6G to
  mmWave: Reduced-Overhead Multi-User Hybrid Beamforming},'' \emph{{IEEE} J.
  Sel. Areas Commun.}, vol.~43, no.~3, pp. 973--987, 2025.

\bibitem{cheng2021hybrid}
Z.~Cheng, Z.~He, and B.~Liao, ``{Hybrid beamforming design for OFDM
  dual-function radar-communication system},'' \emph{{IEEE} J. Sel. Topics
  Signal Process.}, vol.~15, no.~6, pp. 1455--1467, 2021.

\bibitem{li2020dynamic}
H.~Li, M.~Li, Q.~Liu \emph{et~al.}, ``{Dynamic hybrid beamforming with
  low-resolution PSs for wideband mmWave MIMO-OFDM systems},'' \emph{{IEEE} J.
  Sel. Areas Commun.}, vol.~38, no.~9, pp. 2168--2181, 2020.

\bibitem{gardner1986spectral}
W.~A. Gardner, ``{The spectral correlation theory of cyclostationary
  time-series},'' \emph{Signal processing}, vol.~11, no.~1, pp. 13--36, 1986.

\bibitem{liu2022lpi}
X.~Liu, T.~Zhang, X.~Yu, Q.~Shi, G.~Cui, and L.~Kong, ``{LPI waveform design
  for radar system against cyclostationary analysis intercept processing},''
  \emph{Signal Processing}, vol. 201, p. 108681, 2022.

\bibitem{liu2023lpi}
X.~Liu, T.~Zhang, Q.~Shi \emph{et~al.}, ``{LPI radar waveform design with
  desired cyclic spectrum and pulse compression properties},'' \emph{{IEEE}
  Trans. Veh. Technol.}, vol.~72, no.~5, pp. 6789--6793, 2023.

\bibitem{de2008design}
A.~De~Maio, S.~De~Nicola, Y.~Huang \emph{et~al.}, ``{Design of phase codes for
  radar performance optimization with a similarity constraint},'' \emph{{IEEE}
  Trans. Signal Process.}, vol.~57, no.~2, pp. 610--621, 2008.

\bibitem{xu2024enhancing}
L.~Xu, B.~Wang, H.~Li, and Z.~Cheng, ``{Enhancing Physical Layer Security in
  Dual-Function Radar-Communication Systems With Hybrid Beamforming
  Architecture},'' \emph{{IEEE} Wireless Commun. Lett.}, vol.~13, no.~6, pp.
  1566--1570, 2024.

\bibitem{xu2024task}
L.~Xu, B.~Wang, and Z.~Cheng, ``{Task-Oriented Hybrid Beamforming for OFDM-DFRC
  Systems With Flexibly Controlled Space-Frequency Spectra},'' \emph{{IEEE}
  Trans. on Cogn. Commun. Netw.}, vol.~11, no.~1, pp. 375--390, 2025.

\bibitem{yu2016alternating}
X.~Yu, J.-C. Shen, J.~Zhang, and K.~B. Letaief, ``{Alternating minimization
  algorithms for hybrid precoding in millimeter wave MIMO systems},''
  \emph{{IEEE} J. Sel. Areas Commun.}, vol.~10, no.~3, pp. 485--500, 2016.

\end{thebibliography}

	\clearpage

\end{document}